\documentclass[12pt]{article}
\usepackage{natbib,setspace,lscape,longtable}
\usepackage{natbib,epsfig,graphicx}
\usepackage{mathrsfs,amsmath,amsthm,amssymb,color, verbatim}
\usepackage{multirow}

\usepackage{geometry}
\usepackage{float}
\usepackage{array}
\usepackage{lscape}

\usepackage{comment}
\usepackage{setspace}
\usepackage{amsfonts}
\usepackage{indentfirst}
\usepackage{booktabs}
\usepackage{caption}
\usepackage{graphicx, subfig}

\numberwithin{equation}{section}

\newtheorem{theorem}{Theorem}
\newtheorem{lemma}{Lemma}

\newtheorem{proposition}{Proposition}

\newcommand{\csection}[1]
    {\begin{center}
        \stepcounter{section}
        {\bf\large\arabic{section}. #1}
    \end{center}  }
\newcommand{\scsection}[1]
    {\begin{center}
        {\bf\large #1}
    \end{center}
}
\newcommand{\csubsection}[1]{
\begin{center}
\stepcounter{subsection}
{\it\arabic{section}.\arabic{subsection}. #1}
\end{center}
}
\def\1{\mathbf{1}}

\def\mR{\mathbb{R}}

\def\mV{\mathbb{V}}
\def\mX{\mathbb{X}}
\def\mZ{\mathbb{Z}}
\def\mY{\mathbb{Y}}
\def\hat{\widehat}
\def\var{\mbox{var}}

\def\argmax{\mbox{argmax}}

\def\lbk{\left \{ }
\def\rbk{\right \} }

\def\laak{\left \| }
\def\raak{\right \| }

\def\beq{\begin{equation}}
\def\eeq{\end{equation}}
\def\ben{\begin{equation*}}
\def\een{\end{equation*}}
\def\bea{\begin{eqnarray}}
\def\eea{\end{eqnarray}}

\def\bda{\begin{eqnarray*}}
\def\eda{\end{eqnarray*}}

\def\bet{\begin{theorem}}
\def\eet{\end{theorem}}
\def\bel{\begin{lemma}}
\def\eel{\end{lemma}}
\def\bep{\begin{proposition}}
\def\eep{\end{proposition}}
\def\bg{\begin{figure}[tbph]\begin{center}}
\def\eg{\end{center}\end{figure}}

\def\bc{\begin{center}}
\def\ec{\end{center}}

\begin{document}
\begin{center}
{\bf\Large Inward and Outward Network Influence Analysis} \\
\bigskip
Yujia Wu, Wei Lan, Tao Zou and Chih-Ling Tsai \\

{\it Southwestern University of Finance and Economics, The Australian National University and University of California, Davis} \

This Version: \today
\end{center}

\begin{abstract}

Measuring heterogeneous influence across nodes in a network
is critical in network analysis. This paper
proposes an Inward and Outward Network Influence (IONI) model to assess nodal heterogeneity.
Specifically, we allow for two types of influence parameters; one measures the magnitude of influence that each node exerts on others (outward influence),
while we introduce a new parameter
to quantify the receptivity of each node to being influenced by others (inward influence). Accordingly, these two types of influence measures naturally classify
all nodes into four quadrants (high inward and high outward, low inward and high outward,
low inward and low outward, high inward and low outward). To demonstrate our four-quadrant clustering method in practice,
 we apply the quasi-maximum likelihood approach to estimate the influence parameters,
and we show the asymptotic properties of the resulting estimators. In addition,
score tests are proposed
to examine the homogeneity of the two types of influence parameters.
To improve the accuracy of inferences about nodal influences, we introduce
a Bayesian information criterion that selects the optimal
influence model.
The usefulness of the IONI model and the four-quadrant clustering method is illustrated via simulation studies and an empirical example involving customer segmentation.\\


\end{abstract}

\newpage

\csection{INTRODUCTION}

Rapid digital technology advancement has produced large amounts of data that can be used for  network analyses
in various disciplines and professions such as business,
engineering, health care, and science.
For example, Valente (2010) studied social network applications to medical and public health;
Katona et al. (2011) investigated network effects and personal influences in marketing;
Banerjee et al. (2013) examined the diffusion of people's participation in the microfinance program through social networks;
Banerjee et al. (2014) further identified highly central individuals via the diffusion path of the network;
Lawyer (2015) assessed the network influence of spreading processes as seen in epidemic diseases;
Maggio et al. (2019) demonstrated that
the network relationships between brokers and institutional investors can influence market outcomes.
Recently, Tabassum et al. (2018) provided a  succinct overview of social network analysis.

In practice, one of the important tasks in network analysis
 is to identify influential nodes (or actors).
For example, evaluating the effect of each user's behavior on other users or determining the extent of spread from each COVID-19 carrier to other people.
Hence, it is useful to measure the magnitude of the outward influence each  node exerts on others;
see, e.g., Zhu et al. (2019) and Zou et al. (2021).
On the other hand, it is also critical to assess the inward receptivity of each node to being influenced by others.
Examples include analyzing how the behavior of social media influencers is itself
affected by the other users or
measuring the receptivity of each person who is exposed to COVID-19 carriers.
While much research has focused on outward influence, such studies are limited in their predictive power as they neglect the compounding  effect of heterogeneity across nodes in the level of receptivity to influence from others (i.e., inward influence). Our paper improves on prior literature by introducing heterogeneity in  inward influence, in addition to heterogeneity of outward influence. We propose a  model for working with these two influence factors simultaneously. Many existing models that study outward influence alone can be seen as a special case of our model.

To study both types of influences,
we construct a network with $n$
nodes by defining
the adjacency matrix $A=(a_{ij})_{n\times n}\in \mR^{n\times n}$ to characterize the relationship between any two adjacent nodes in the network
(see, e.g., Carrington et al. 2005 and Scott 2013). Specifically, we define
$a_{ij}=1$ if there is a direct connection from node $i$ to node $j$ and $a_{ij}=0$ otherwise. In addition, we denote $a_{ii}=0$ for any $1\leq i\leq n$.
For example, we define $a_{ij}=1$ if user $i$ in the Sina Weibo network follows user $j$  and $a_{ij}=0$ otherwise; see the real data analysis in Section 3.2.
Subsequently, we normalize $a_{ij}$ and obtain $W=(w_{ij})\in\mR^{n\times n}$ with $w_{ij}=a_{ij}/\sum_{j=1}^{n}a_{ij}$, which is the weighted adjacency matrix.

After constructing the network structure, we then define the response variable $Y=(Y_1, \cdots, Y_n)^\top$ that usually depends on the target of the study.
For example,
suppose that our aim is to
 identify the specific users who can most influence others' activity
in the Sina Weibo network given in Section 3.2.
The $Y_i$ ($i=1,\cdots,n$) could be the number of posts made by the $i$-th user as a measure of user $i$'s activity.
Although the response of the $i$-th node, $Y_i$, can be influenced by the other nodes' responses,
it can also depend on its own attributes, such as gender and duration in the Sina Weibo network, and we name this vector of attributes  the nodal covariate $X_{i}$.
Therefore, to identify influential nodes, for each node $1\leq i\leq n$, we  consider the continuous response variable $Y_{i}\in \mR^1$ and its associated $p$-dimensional nodal covariate $X_{i}=(X_{i1},\cdots,X_{ip})^\top\in \mR^p$. By integrating this nodal information with the network structure induced by the adjacency matrix $A$, we propose the following model
\begin{equation}
    Y_{i}=\sum\limits_{j=1}^{n} b_{ij}Y_{j}+X_{i}^\top\eta+\varepsilon_{i},\label{equ:b}
\end{equation}
where $b_{ij}$ is the function of $w_{ij}$ that characterizes the influence effect between nodes $i$ and $j$ for $i, j=1,\cdots, n$,
$\eta=(\eta_{1},\cdots,\eta_{p})^\top \in \mR^p$ is the $p$-dimensional unknown regression coefficient vector, and $\varepsilon_{i}$ is the random error for $i=1,\cdots, n$.
If we parameterize $b_{ij}=\lambda w_{ij}$ by a single influence parameter $\lambda$,
then the resulting autoregressive model has been studied for a long history in the field of
economics (see, e.g., Anselin 1988, Manski 1993a, Manski 1993b, Bramoull\'e et al. 2009, LeSage and Pace 2009, Liu and Lee 2010, Gupta and Robinson 2015, Zhou et al. 2017, Zhang and Yu 2018, Gao et al. 2019, Kwok 2019 and Zhu et al. 2019). It is of interest to note that this type of model was labeled as
``mixed-regression", ``spatial autoregressive'', ``spatial correlation'', and ``spatial autoregressive model (SAR)'' by Anselin (1988), Manski (1993a), and  LeSage and Pace (2009), respectively, although the term SAR is most commonly used  in  the literature.

As Gupta and Robinson (2015) and
Lam and Souza (2019) noted, using the single parameter $\lambda$ to characterize the influence
effect among the $n$ nodes may not capture all of their heterogeneity.
To alleviate this limitation, Zhu et al. (2019) and Zou et al. (2021) proposed the network influence model by parameterizing
$b_{ij}$ in (\ref{equ:b}) as $\lambda_j w_{ij}$.
Thus, the influence effect of node $j$ on node $i$ not only relies on the network structure $w_{ij}$,
but also depends on $\lambda_j$. We name $\lambda_j$ the outward influence index of node $j$.
It is worth noting that the influence of node $j$ on node $i$ does not take into account the magnitude of the response of node $i$. Hence, we cannot measure the true influence of node $j$ accurately.
For example, consider a Sina Weibo network with three nodes, $j_1$, $j_2$ and $j_3$. Suppose that $j_1$ is a famous star
and both nodes $j_2$ and $j_3$ follow node $j_1$, while  node $j_2$ is quite active  and node $j_3$ is inactive. Under this scenario, we expect that $j_1$ has larger influence
on $j_2$ than on $j_3$ since $j_2$ renders   more responses  to $j_1$.
This situation can also occur in the infectious diseases network. For instance, let
$j_1$ be a super spreader, $j_2$ be an older person with chronic susceptibility, and $j_3$ be a younger person with a strong immune system. It is expected that $j_1$ will have more apparent influence on the health of $j_2$ than on the health of $j_3$ since $j_2$ is more likely than $j_3$ to yield an immune response  to $j_1$.
Thus, we cannot detect potential collective outcomes accurately by considering the outward influence index alone.
This motivates us to introduce an inward influence measure
that can characterize each individual's own response
 to others' influence.

To determine the real influence, we introduce the Inward and Outward Network Influence (IONI) model given below by assuming  $b_{ij}$ in (\ref{equ:b}) to be $\gamma_i w_{ij} \lambda_j$.
\begin{equation}
    Y_{i}=\sum\limits_{j=1}^{n}\big(\gamma_i w_{ij}\lambda_{j}\big)Y_{j}+X_{i}^\top\eta+\varepsilon_{i},\label{equ:NIFA}
\end{equation}
where $\gamma_i$ measures the degree of node $i$'s receptivity to being influenced by others and
$\lambda_j$ determines the influence magnitude of node $j$ on others.
Hence, we name
$\gamma_i$ the inward influence index,
and $\lambda_j$ is the outward influence index mentioned earlier.
In addition, we assume that $0\le\gamma_i\le 1$
and $0\le \lambda_j\le 1$.
It is worth noting that
the influence indices $\gamma_i$ and $\lambda_j$ are usually related to the nodes' attributes (see, e.g., Trusov et al. 2010 and Zou et al. 2021), which we denote them
 $Z_i$ and $V_j$, respectively.
For instance, in the Sina Weibo example, $Z_i$ and $V_j$ can include the number of each user's followers and the number of each  user's followees, i.e.,
other people that a given user is following.
Accordingly,
we  parameterize $\gamma_i$ and $\lambda_j$ as the known functions of the nodes' attributes $Z_i$ and $V_j$, respectively, in Section 2.1.

To illustrate the usefulness of the  $\gamma$ and $\lambda$ indices, we consider customer segmentation in marketing as an application.
Figure 1 depicts $\gamma$ and $\lambda$, which
increase in the directions of the $x$ and $y$ arrows, respectively, with the origin $(\gamma_0, \lambda_0)$.
Then  we classify the nodes of the network into four groups according to the values of $\gamma$ and $\lambda$.
The first  quadrant of Figure 1 is group I, which consists of customers
with high $\gamma$ and $\lambda$.
Accordingly, those customers not only have strong influence over others, but also can be influenced greatly by others.
This type of customer shares similar properties to ``early adopters" in marketing; see, e.g., Wolf and Seebauer (2014), Catalini and Tucker (2017).
Hence, identifying this type of customer is essential for promoting new products.
The second quadrant is group II, in which the customers have low $\gamma$ and high $\lambda$. This group's customers tend to influence others, but not vice versa. They can be
named ``opinion leaders" or ``social influencers;" see, e.g.,
Li et al. (2013), Ma and Liu (2014) and Bamakan et al. (2019).
Group III is located in the third quadrant, in which the customers have both low $\gamma$ and  $\lambda$.
This type of customer neither exerts influence nor is influenced by others. Hence, these customers do not create awareness or show interest in products, and they can be considered  ``inactive customers;" see Kumar and Reinartz (2018).
Finally, group IV is located in the fourth quadrant, in  which the customers
have high $\gamma$ and low $\lambda$.
This type of customer seems to be very susceptible but has little influence on others, and they can be treated as
the ``early majority;" see, e.g., Ram and Jung (1994) and Mattila et al. (2003).

Based on the above four-quadrant clustering, practitioners can segment Sina Weibo's users
into the four groups to improve product promotion or opinion dissemination.
For example,
 if practitioners need to promote a new product on Sina Weibo, they could target
the ``early adopters" (i.e., the users with high $\gamma$ and high $\lambda$). In contrast, if practitioners
would like to extend the life cycle of mature products, they should focus on the ``early
majority" (i.e., the users with high $\gamma$ and low $\lambda$). To effectively make use of the above two
types of customers, decision makers should focus on the ``opinion leaders" (i.e., the users
with low $\gamma$ and high $\lambda$) for  promotions. Finally, removing or ignoring prolonged
``inactive users" (i.e., the users with low $\gamma$ and low $\lambda$) can cut down on promoting costs. We provide detailed illustrations in Section 3.2.

\begin{figure}[H]
 \centering
 \includegraphics[width=3in,height=2in]{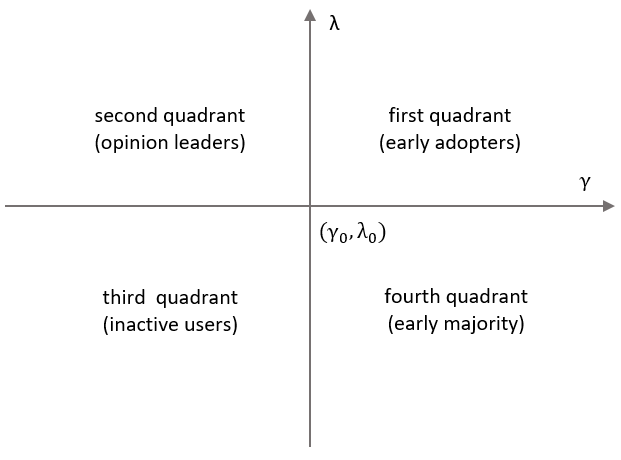}
 \caption{ Location of customer segmentation }
 \label{fig1}
\end{figure}

The aim of this paper is to propose the Inward and Outward Network Influence (IONI) model, which allows us to identify four types of influencers in network analysis.
To estimate the model and influence indices, we study identification conditions and
parameterize $\lambda$ and $\gamma$ as known functions of the nodes' attributes, respectively.
 Then, we employ Lee's (2004) quasi-maximum likelihood approach to obtain parameter estimators.
In addition, we demonstrate their asymptotic properties without imposing any specific error distributions. A Bayesian information criterion is also proposed
for selecting optimal models, and its consistency is demonstrated. Moreover, score tests are provided to examine the homogeneity of $\lambda$ and $\gamma$.
Simulation studies and an empirical example for differentiating the four different types of users in the Sina
Weibo network are presented to
demonstrate the utility of the proposed model.
The rest of the article is organized as follows. Section 2
describes the model structure in the matrix form, parameter estimators,
four-quadrant clustering, selection criterion and test statistics. Section 3 presents numerical studies. We provide a discussion
with some concluding remarks in Section 4. All technical details are relegated to the supplementary material.

\csection{METHODOLOGY AND THEORETICAL RESULTS}

\csubsection{Model Setting and Identification Condition}

To effectively examine the Inward and Outward Network Influence (IONI) model, we express the proposed model (\ref{equ:NIFA}) in the matrix form as follows:
\begin{equation}
\mathbb{Y}=\Gamma W\Lambda\mathbb{Y}+\mathbb{X}\eta+\varepsilon,\label{equ:matrix}
\end{equation}
where $\mathbb{Y}=(Y_1, \cdots, Y_n)^\top\in\mR^n$, $\mathbb{X}=(X_1^\top, \cdots, X_n^\top)^\top\in\mR^{n\times p}$,
$\Gamma =\mbox{diag}\{\gamma_{1},\cdots,\gamma_{n}\}$, $\Lambda =\mbox{diag}\{\lambda_{1},\cdots,\lambda_{n}\}$, $\varepsilon=(\varepsilon_1, \cdots, \varepsilon_n)^\top\in\mR^n$,
and $\varepsilon_i$ are independent and identically distributed with mean 0 and variance $\sigma^2$ for
$i=1, \cdots, n$.
In addition, $\gamma_{i}$ (and $\lambda_i$) are the inward (and outward)  influence parameters or indices, respectively.

It is worth noting that there are $2n$ unknown parameters in (\ref{equ:matrix}),
which cannot be estimated via the $n$ observations.  However,  $\gamma_i$ and $\lambda_j$ can each be considered a function of the associated attributes of nodes $i$ and $j$, respectively.
To this end,
we parameterize $\gamma_{i}$ with $\gamma_{i}(\alpha)=F(Z^{\top}_{i}\alpha)$ and $\lambda_j$ with $\lambda_j(\beta) = F(V^{\top}_j\beta)$,  where $F(\cdot)$ is a
strictly monotone and known function,
and $Z_i$ and $V_j$ are the corresponding attributes associated
with the inward and outward influence indices, $\gamma_{i}$ and $\lambda_j$, respectively.
In addition,
$Z_{i} = (z_{i1},\cdots,z_{i{d_1}})^{\top}\in \mR^{d_{1}\times1}$  with $z_{i1}\equiv 1$ and $V_j= (v_{j1},\cdots,v_{jd_{2}})^{\top}\in \mR^{d_{2}\times1}$ with $v_{j1}\equiv 1$,
and the corresponding  unknown regression coefficient vectors are
$\alpha=(\alpha_{1},\cdots,\alpha_{d_{1}})^{\top}\in \mR^{d_{1}\times1}$
and $\beta=(\beta_{1},\cdots,\beta_{d_{2}})^{\top}\in \mR^{d_{2}\times1}$.
Moreover, define $Z_{-1,i}=(z_{i2},\cdots,z_{i{d_1}})^{\top}$ and
$V_{-1, j}=(v_{j2},\cdots,v_{jd_{2}})^{\top}$, and assume that $\mZ_{-1}=(Z_{-1,1}, \cdots, Z_{-1, n})^\top$ and
$\mV_{-1}=(V_{-1,1}, \cdots, V_{-1, n})^\top$ are of full rank.
Since both $\Lambda$ and $\Gamma$ can be expressed as a function of $\alpha$ and $\beta$, respectively, we express (\ref{equ:matrix}) as:
\begin{equation}
\mathbb{Y}=\Gamma(\alpha)W\Lambda(\beta)\mathbb{Y}+\mathbb{X}\eta+\varepsilon.\label{equ:paramfun}
\end{equation}
To make the proposed model (\ref{equ:paramfun}) practically useful, one needs to specify
the function $F(\cdot)$ that links attributes to influence parameters.
We first adopt the sigmoid function, which is often used in machine learning (see, e.g.,  Hastie et al, 2009). That is,
$F(Z^{\top}_{i}\alpha)=e^{Z_{i}^\top\alpha}/(1+e^{Z_{i}^\top\alpha})$. We next
consider two related link functions: the inverse of log-log (i.e., $F(Z_{i}^\top\alpha)=1-e^{-e^{Z_{i}^\top\alpha}})$ and the inverse of probit (i.e., $F(Z_{i}^\top\alpha)=\Phi(Z_{i}^\top\alpha)$), where $\Phi(\cdot)$ is the distribution function of the standard normal distribution.

In order to estimate the unknown parameters in model (\ref{equ:paramfun}), one  needs to consider the identification problem of  $\gamma_i$ and $\lambda_j$.
That is, for any $\Gamma =\mbox{diag}\{\gamma_{1},\cdots,\gamma_{n}\}$, $\Lambda =\mbox{diag}\{\lambda_{1},\cdots,\lambda_{n}\}$,
$\widetilde{\Gamma} =\mbox{diag}\{\widetilde \gamma_{1},\cdots,\widetilde \gamma_{n}\}$ and $\widetilde\Lambda =\mbox{diag}\{\widetilde\lambda_{1},\cdots,\widetilde\lambda_{n}\}$,
$\Gamma W\Lambda=\widetilde{\Gamma}W\widetilde\Lambda$ leads to $\widetilde{\Gamma}=\Gamma$ and $\widetilde{\Lambda}=\Lambda$, where
$\gamma_{i}=\gamma_{i}(\alpha)=F(Z^{\top}_{i}\alpha)$, $\lambda_j=\lambda_j(\beta)= F(V^{\top}_j\beta)$, $\widetilde \gamma_{i}=\widetilde \gamma_{i}(\widetilde \alpha)=F(Z^{\top}_{i}\widetilde\alpha)$,
$\widetilde\lambda_j=\widetilde\lambda_j(\widetilde\beta) = F(V^{\top}_j\widetilde\beta)$,
$\widetilde\alpha=(\widetilde\alpha_{1},\cdots,\widetilde\alpha_{d_{1}})^{\top}$
and $\widetilde\beta=(\widetilde\beta_{1},\cdots,\widetilde\beta_{d_{2}})^{\top}$.
Specifically,
if $\Gamma W\Lambda=\widetilde{\Gamma}W\widetilde\Lambda$ and $w_{ij}>0$
for some $i\ne j$, then
$F(Z^{\top}_{i}\alpha)/F(Z^{\top}_i\widetilde\alpha)=F(V^{\top}_j\widetilde\beta)/F(V^{\top}_j\beta)=c$ for some constant $c$.
To this end, we impose the constraint that either $\alpha_1=0$ or $\beta_1=0$, to assure identifiability. To illustrate the necessity for this constraint,
we define $\alpha_{-1}=(\alpha_2, \cdots, \alpha_{d_1})^\top\in\mR^{d_1-1}$
and $\beta_{-1}=(\beta_2, \cdots, \beta_{d_1})^\top\in\mR^{d_2-1}$.
Suppose that $\alpha_{-1}=0$ and $\beta_{-1}=0$. Then $F(Z^{\top}_{i}\alpha)/F(Z^{\top}_i\widetilde\alpha)=c$
implies that $\widetilde\alpha_{-1}=0$ and $F(\alpha_1)/F(\widetilde\alpha_1)=c$. Similarly, we obtain $\widetilde\beta_{-1}=0$ and $F(\widetilde\beta_1)/F(\beta_1)=c$.
Thus, $F(\alpha_1)F(\beta_1)=F(\widetilde\alpha_1)F(\widetilde\beta_1)$.
Using the fact that $F(\cdot)$ is
strictly monotone, the model cannot be identified unless we impose the constraint that
either $\alpha_1=0$ or $\beta_1=0$.

Without loss
of generality, we assume $\alpha_1=0$, or equivalently $Z_i^\top\alpha=Z_{-1,i}^\top\alpha_{-1}$. For the sake of simplification, however,  we
slightly abuse notation and still use
$\alpha=(\alpha_{1},\cdots,\alpha_{d_{1}})^{\top}\in \mR^{d_{1}\times1}$
and $Z_{i} = (z_{i1},\cdots,z_{i{d_1}})^{\top}\in \mR^{d_{1}\times1}$ hereafter,
rather than
$\alpha_{-1}$ and $Z_{-1,i}$,  as the parameter vector and its associated attributes, respectively.

\noindent\textbf{Remark 1:}
In network analysis,  most existing studies have focused on outward influence.
To construct  spatio-temporal models for panel
data, however, Dou et al. (2016) introduced a pure spatial effect measure.
Without considering the temporal structure,
we can consider
 their measure to be inward influence
by parameterizing $b_{ij}$ as $\gamma_i w_{ij}$.
Dou et al. (2016) allowed the temporal observations to tend toward infinity and showed that their parameters are estimable without introducing a link function.
However, our model does not consider the temporal structure. Hence, we introduce link functions for influence parameter estimation.

\csubsection{Parameter Estimation}

In  model (\ref{equ:paramfun}), we do not assume that the random errors  are normally distributed. To obtain nice properties of the  estimators,
we adopt Lee's (2004) approach by employing the quasi-maximum likelihood estimation (QMLE) method to estimate unknown parameters.
Define $S(\alpha,\beta)= I_{n}-\Gamma(\alpha)W\Lambda(\beta)$. We then have
$\varepsilon=\varepsilon(\eta,\alpha,\beta)=S(\alpha,\beta)\mY-\mathbb{X}\eta$,
and the normal log-likelihood function of (\ref{equ:paramfun}) is
\begin{equation}
\ell(\theta)=-\frac{n}{2}\log 2\pi-\frac{n}{2}\log\sigma^{2}-\frac{1}{2\sigma^{2}}(S(\alpha,\beta)\mathbb{Y}-\mathbb{X}\eta)^\top(S(\alpha,\beta)\mathbb{Y}-\mathbb{X}\eta)+\log|\det\{S(\alpha,\beta)\}|,\nonumber
\end{equation}
where $\theta=(\eta^\top,\alpha^\top,\beta^\top,\sigma^{2})^\top$.
We next apply the concentrated QMLE method to estimate the parameters. Specifically, given $\alpha$ and $\beta$,
we maximize $\ell(\theta)$ with respect to $\eta$ and $\sigma^2$, which leads to
\begin{equation}
\hat{\eta}(\alpha,\beta)=(\mathbb{X}^\top\mathbb{X})^{-1}\mathbb{X}^\top S(\alpha,\beta)\mathbb{Y} ~~~{\mbox{and}}\nonumber
\end{equation}
\begin{equation}
\hat{\sigma}^{2}(\hat{\eta}(\alpha,\beta),\alpha,\beta)=n^{-1}\varepsilon^\top(\hat{\eta}(\alpha,\beta),\alpha,\beta)
\varepsilon(\hat{\eta}(\alpha,\beta),\alpha,\beta), \nonumber
\end{equation}
where $\varepsilon(\hat{\eta}(\alpha,\beta),\alpha,\beta)=\mathcal{M}_{\mathbb{X}}S(\alpha,\beta)\mathbb{Y}$ and $\mathcal{M}_{\mathbb{X}} = I_{n}-\mathbb{X}(\mathbb{X}^\top\mathbb{X})^{-1}\mathbb{X}^\top$. The resulting concentrated quasi-log-likelihood is
\begin{equation}
\ell_{c}(\alpha,\beta)=-\frac{n}{2}\log2\pi-\frac{n}{2}-\frac{n}{2}\log\hat{\sigma}^{2}(\hat{\eta}(\alpha,\beta),\alpha,\beta)+\log|\det\{S(\alpha,\beta)\}|.\label{equ:quasi-log-likelihood}
\end{equation}
By employing the most commonly used quasi-Newton method to maximize the above equation with respect to $\alpha$ and $\beta$, we get the  QMLE, $(\hat{\alpha}^\top,\hat{\beta}^\top)^\top$= $\argmax\ell_{c}(\alpha,\beta)$.
We then obtain the QMLEs of $\eta$ and $\sigma^{2}$, which are $\hat{\eta}=\hat{\eta}(\hat\alpha,\hat\beta)$ and $\hat{\sigma}^{2}=\hat{\sigma}^{2}(\hat{\eta},\hat\alpha,\hat\beta)$.
To develop the asymptotic distribution of estimated parameters, we introduce some notation and equations  below.

Let $\Gamma_{\alpha_{k}}(\alpha):=\partial\Gamma(\alpha)/\partial\alpha_{k}=\mbox{diag}\{z_{1k}F'(Z_{1}^\top \alpha),\cdots,z_{nk}F'(Z_{n}^\top \alpha)\}$
for $k = 1,\cdots,d_{1}$,   and $\Lambda_{\beta_{l}}(\beta):=\partial\Lambda(\beta)/\partial\beta_{l}=\mbox{diag}\{v_{1l}F'(V_{1}^\top {\beta}),\cdots,v_{nl}F'(V_{n}^\top \beta)\}$ for $l = 1,\cdots,d_2$, where $F'(\cdot)$ is the first derivative of $F$.
In addition, we denote the
Fisher information matrix of the quasi-maximum likelihood $l(\theta)$ by
$\mathcal{I}_{n}(\theta)=-\frac{1}{n}E\big[\frac{\partial^{2}\ell(\theta)}{\partial\theta\partial\theta^\top }\big]$.
To show the asymptotic properties of parameter estimators,
we define $\mathcal{J}_{n}(\theta,\mu_{3},\mu_{4})=\frac{1}{n}\var\big\{\frac{\partial\ell(\theta)}{\partial\theta} \big\}-\mathcal{I}_{n}(\theta)$,
where the third-order and fourth-order moments are $\mu_{3} =
E(\epsilon_{i}^{3})$ and  $\mu_{4} =E(\epsilon_{i}^{4})$, respectively.
The detailed calculations of $\mathcal{I}_{n}(\theta)$ and $\mathcal{J}_{n}(\theta,\mu_{3},\mu_{4})$ are given in Section 1 of the supplementary material.
To further obtain the theoretical properties of $\hat{\gamma}$ and $\hat{\lambda}$,
we define $l_\alpha=(0_{d_1\times p},I_{d_1}, 0_{d_1\times (d_2+1)})\in\mR^{d_1\times (p+d_1+d_2+1)}$ and
$l_\beta=(0_{d_2\times (p+d_1)},I_{d_2}, 0_{d_2\times 1})\in\mR^{d_2\times (p+d_1+d_2+1)}$, where $0_{K_1\times K_2}$ denotes a $K_1\times K_2$ zero matrix,
and denote
\[ \mathcal{D}_{\alpha,n}(\theta,\mu_{3},\mu_{4})=
 l_\alpha\Big\{\mathcal{I}_n^{-1}(\theta)+\mathcal{I}_n^{-1}(\theta)\mathcal{J}_n(\theta,\mu_{3},\mu_{4})\mathcal{I}_n^{-1}(\theta)\Big\}l_\alpha^\top \mbox{~and~}\]
 \beq \mathcal{D}_{\beta,n}(\theta,\mu_{3},\mu_{4})=
 l_\beta\Big\{\mathcal{I}_n^{-1}(\theta)+\mathcal{I}_n^{-1}(\theta)\mathcal{J}_n(\theta,\mu_{3},\mu_{4})\mathcal{I}_n^{-1}(\theta)\Big\}l_\beta^\top. \eeq

We next define the necessary norms and introduce conditions used in the theoretical proofs.
Denote $\|\cdot\|_s$ the $s$-norm of a vector (or a matrix)
for $1\leq s\leq \infty$. Specifically, for any generic vector
$x=(x_1,\cdots,x_q)^\top\in\mathbb{R}^q$, $\|x\|_s=(\sum_{i=1}^q |x_i|^s)^{1/s}$, and, for any generic matrix $G\in\mathbb{R}^{m\times q}$,
\[
\laak G\raak_s=\sup\lbk \frac{\laak Gx\raak_s}{\laak x\raak_s}:x\in\mathbb{R}^{q\times 1}\textrm{ and }x\neq 0\rbk.
\]
In addition, define the element-wise $\ell_{\infty}$ norm for any generic matrix $G$ as $|G|_{\infty}=\|\textrm{vec}(G)\|_{\infty}$, where $\textrm{vec}(G)$ denotes the vectorization for any generic matrix $G$. We subsequently present five technical conditions, which are required to  ensure the theoretical properties of the estimates.

\begin{itemize}
\item [(C1)] Assume that the random errors $\varepsilon_{i}$ are independent and identically distributed with mean 0, and $E|\varepsilon_{i}|^{4+\xi}<\infty$ for some $\xi > 0$.
\item [(C2)] Assume $\sup_{n\ge1}||W||_1<\infty$ and $\sup_{n\ge1}||W||_{\infty}<\infty$.
\item [(C3)] Assume that $S(\alpha,\beta)= I_{n}-\Gamma(\alpha)W\Lambda(\beta)$ is nonsingular uniformly over $\alpha$ and $\beta$ in the compact parameter space $\mathcal{P}$,
and the true values of $\alpha$ and $\beta$ are also in the interior of $\mathcal{P}$. In addition, assume that $\sup_{\{\alpha,\beta\}\in\mathcal{P}}\sup_{n\ge1}||S^{-1}(\alpha,\beta)||_{1}<\infty$ and $\sup_{\{\alpha,\beta\}\in\mathcal{P}}\sup_{n\ge1}||S^{-1}(\alpha,\beta)||_{\infty}<\infty$ hold.
\item [(C4)] Assume $\sup_{n\ge1}|\mX|_{\infty}< \infty$. In addition, for the true parameters $\alpha$ and $\beta$, assume that
\begin{center}
    $\mathop{\sup}\limits_{n\ge1}\mathop{\max}\limits_{1\leq i\leq n}|z_{ik_{1}}F'(Z_{i}^\top \alpha)|<\infty$, \quad $\mathop{\sup}\limits_{n\ge1}\mathop{\max}\limits_{1\leq i\leq n}|v_{il_{1}}F'(V_{i}^\top \beta)|<\infty$,~\\
    ~\\
    $\mathop{\sup}\limits_{n\ge1}\mathop{\max}\limits_{1\leq i\leq n}|z_{ik_{1}}z_{ik_{2}}F''(Z_{i}^\top \alpha)|<\infty$, \quad $\mathop{\sup}\limits_{n\ge1}\mathop{\max}\limits_{1\leq i\leq n}|v_{il_{1}}v_{il_{2}}F''(V_{i}^\top \beta)|<\infty,$~\\
    ~\\
    $\mathop{\sup}\limits_{n\ge1}\mathop{\max}\limits_{1\leq i\leq n}|z_{ik_{1}}z_{ik_{2}}z_{ik_{3}}F'''(Z_{i}^\top \alpha)|<\infty$, \quad $\mathop{\sup}\limits_{n\ge1}\mathop{\max}\limits_{1\leq i\leq n}|v_{il_{1}}v_{il_{2}}v_{il_{3}}F'''(V_{i}^\top \beta)|<\infty,$
\end{center}
for any $k_{1},k_{2},k_{3}\in \{1,\cdots,d_{1}\}$ and $l_{1},l_{2},l_{3}\in \{1,\cdots,d_{2}\}$, where the link function $F$ is assumed to be three times differentiable.
\item [(C5)] Assume $\mathcal{I}_{n}(\theta)\to\mathcal{I}(\theta)$ and $\mathcal{J}_{n}(\theta,\mu_{3},\mu_{4})\to\mathcal{J}(\theta,\mu_{3},\mu_{4})$ as $n\to\infty$. We further assume that $\mathcal{I}(\theta)$ and $\mathcal{I}(\theta)+\mathcal{J}(\theta,\mu_{3},\mu_{4})$ are finite and positive definite, where $\mathcal{I}_{n}(\theta)$ and $\mathcal{J}_{n}(\theta,\mu_{3},\mu_{4})$
    are defined in  Section 1 of the supplementary material.
\end{itemize}

Condition (C1) is a moment condition, which means it is weaker than a distribution assumption,
such as the normal distribution considered in Zhou et al. (2017) and sub-Gaussian assumption considered in Zhu et al. (2019).
Conditions (C2) and (C3) are standard regularity conditions used in spatial
literature (Lee 2004 and Yu et al. 2008). We can verify that these two conditions are
satisfied as long as the network is sparse so that
each individual node is  connected only to a finite number of other nodes
(i.e., $\sum_{j=1}^na_{ij}<\infty$ for any $i=1,\cdots,n$).
Accordingly, $\sup_{n\ge1}||W||_{\infty}<\infty$ and $\sup_{n\ge1}||W||_1<\infty$ are  satisfied.
In addition, $1\le \sum_{j=1}^n|\gamma_i\lambda_jw_{ij}|+1<\infty$ for any $i$.
This implies that
$1\le\inf\{\frac{\|S(\alpha,\beta)x\|_\infty}{\|x\|_\infty}\}<\infty$.
By the definition of
  $\|S^{-1}(\alpha,\beta)\|_s=1/\inf\{\frac{\|S(\alpha,\beta)x\|_s}{\|x\|_s}:x\in \mR^{n}\}$, we immediately obtain
  $\sup_{\{\alpha,\beta\}\in\mathcal{P}}\sup_{n\ge1}||S^{-1}(\alpha,\beta)||_{\infty}<\infty$.
Employing a similar approach, $\sup_{\{\alpha,\beta\}\in\mathcal{P}}\sup_{n\ge1}||S^{-1}(\alpha,\beta)||_1<\infty$ can be established.

Condition (C4) imposes conditions on the link function and attributes,  and it can be satisfied as long as
the link function $F(\cdot)$ has bounded first  three derivatives and the elements of attributes $X_i$, $Z_i$ and $V_i$ are uniformly bounded constants for all
$i$ (see, e.g., Assumption 6 in Lee 2004). It is worth noting that the first three derivatives of the three link functions (i.e., sigmoid function,  inverse of log-log, and inverse of probit)
mentioned in Section 2.1 are bounded and they satisfy this condition.
Condition (C5) is a type of law of large numbers assumption for ensuring the convergences of the Hessian matrix and the variance of the
score function, in order to establish the asymptotic normality of $\hat{\theta}$. Similar conditions can be found in Fan and Li (2001), Lee (2004) and Zhu et al. (2017). Under Condition (C5),
 the concentrated quasi-log-likelihood function in (\ref{equ:quasi-log-likelihood}) is concave and has a global maximizer.
Based on the above notations and conditions, we then have the following results.
\bet
Under Conditions (C1)-(C5), we then have (i) $\sqrt{n}(\hat{\theta}-\theta)$ is
 asymptotically distributed as $N(0,\mathcal{I}^{-1}(\theta)+\mathcal{I}^{-1}(\theta)\mathcal{J}(\theta,\mu_{3},\mu_{4})\mathcal{I}^{-1}(\theta))$; and
 (ii) for any fixed $i\geq 1$,
 $\sqrt{n}(\hat{\gamma}_i-\gamma_i)\stackrel{d}{\longrightarrow}N\big(0,\{F'(Z_i^\top\alpha)\}^2Z_i^\top\mathcal{D}_\alpha(\theta,\mu_{3},\mu_{4})Z_i\big)$
 and $\sqrt{n}(\hat{\lambda}_i-\lambda_i)\stackrel{d}{\longrightarrow} N\big(0,\{F'(V_i^\top\beta)\}^2V_i^\top\mathcal{D}_\beta(\theta,\mu_{3},\mu_{4})V_i\big)$,
 where $\mathcal{I}(\theta)$ and $\mathcal{J}(\theta,\mu_{3},\mu_{4})$ are defined
in Condition (C5), and
$\mathcal{D}_\alpha(\theta,\mu_{3},\mu_{4})$ and $\mathcal{D}_\beta(\theta,\mu_{3},\mu_{4})$ are the correspondingly convergent matrices of $\mathcal{D}_{\alpha,n}(\theta,\mu_{3},\mu_{4})$ and $\mathcal{D}_{\beta,n}(\theta,\mu_{3},\mu_{4})$ defined  above.
\eet

In practice, we can estimate the unknown quantities $\mathcal{D}_\alpha(\theta,\mu_{3},\mu_{4})$
and $\mathcal{D}_\beta(\theta,\mu_{3},\mu_{4})$ consistently by their corresponding estimators $\mathcal{D}_{\alpha,n}(\hat{\theta},\hat{\mu}_{3},\hat{\mu}_{4})$ and
 $\mathcal{D}_{\beta,n}(\hat{\theta},\hat{\mu}_{3},\hat{\mu}_{4})$, where
$\hat\mu_{3}=n^{-1}\sum_i \hat\varepsilon_i^3$, $\hat\mu_{4}=n^{-1}\sum_i \hat\varepsilon_i^4$, and
$\hat\varepsilon=\varepsilon(\hat\eta,\hat\alpha,\hat\beta)=(\hat\varepsilon_1,\cdots, \hat\varepsilon_n)^\top\in\mR^n$.
Using the continuous mapping theorem, we can also estimate $F'(Z_i^\top\alpha)$ and $F'(V_i^\top\beta)$ consistently by the estimators $F'(Z_i^\top\hat{\alpha})$ and $F'(V_i^\top\hat{\beta})$, respectively.

\csubsection{Four-Quadrant Clustering}

To make use of the inward and outward influence indices, we can apply the four-quadrant clustering approach to
partition all nodes into four groups with known threshold values $\gamma_0$ and $\lambda_0$ as depicted in Figure 1.
To achieve this end, a heuristic approach can be applied to select the threshold values $\gamma_0$ and $\lambda_0$ via the empirical quantiles and then classify nodes by
comparing their estimated inward and outward influence indices with the threshold values, respectively.
However, this simple heuristic approach does not take into account the variation in the
    estimated inward and outward influence indices. To this end, we consider the following two hypotheses for $i=1,\cdots,n$:
\begin{center}
$H_{0i,\gamma}$: $\gamma_{i}\leq \gamma_0$ \mbox{~vs.~} $H_{1i,\gamma}: \gamma_{i}> \gamma_0$  and~\\
$H_{0i,\lambda}$: $\lambda_{i}\leq \lambda_0$ \mbox{~vs.~} $H_{1i,\lambda}: \lambda_{i}> \lambda_0$.
\end{center}
Their corresponding test statistics are
\begin{center}
$T_{i, \gamma}=\frac{\sqrt{n}(\hat{\gamma}_{i}-\gamma_0)}{\sqrt{\{F'(Z_{i}^\top\hat{\alpha})\}^2Z_{i}^\top\mathcal{D}_{\alpha,n}(\hat\theta,\hat\mu_{3},\hat\mu_{4})Z_{i}}}$ and
$T_{i, \lambda}=\frac{\sqrt{n}(\hat{\lambda}_{i}-\lambda_0)}{\sqrt{\{F'(V_{i}^\top\hat{\beta})\}^2V_{i}^\top\mathcal{D}_{\beta,n}(\hat\theta,\hat\mu_{3},\hat\mu_{4})V_{i}}}$,
\end{center}
where $\mathcal{D}_{\alpha,n}(\hat{\theta},\hat{\mu}_3,\hat{\mu}_4)$ and $\mathcal{D}_{\beta,n}(\hat{\theta},\hat{\mu}_3,\hat{\mu}_4)$ are given above in the discussion of Theorem 1.
By Theorem 1, we can immediately obtain that
$T_{i, \gamma}$ and $T_{i, \lambda}$
are asymptotically distributed as $N(0,1)$, which yields their respective
$p$-values.

It is worth noting that the above testing procedure implicitly involves $n$ multiple tests. This motivates us to control the false discovery rate
(FDR) by selecting the threshold of the $p$-values obtained from the multiple tests and then identifying significant hypotheses and determining the rejection regions (see, e.g., Storey et al. 2004).

In sum, Theorem 1 not only provides the asymptotic properties of parameter estimators, but also allows us to classify existing nodes into adequate categories via inward and outward influence indices. Thus, the four-quadrant clustering method can play an important role in network applications.

\csubsection{Variable Selections}

The influence parameters $\gamma_i$ and $\lambda_j$ in  model (2.2) are  known functions of attributes. To make better inferences and predictions, we employ Schwarz's Bayesian information criterion to select attributes in the correct model consistently.
Define the full model $S_{F}=\{S_{F\alpha},S_{F\beta}\}$, where $S_{F\alpha}=\{1,\cdots,d_{1}\}$ and $S_{F\beta}=\{1,\cdots,d_{2}\}$.
In addition, denote true model as $S_T =\{S_{T\alpha},S_{T\beta}\}$,
where $S_{T\alpha}=\{i\ge1, \alpha_{i}\ne0\}$ and $S_{T\beta}=\{j\ge1, \beta_{j}\ne0\}$.
Accordingly, the full model includes all possible attributes (i.e., covariates), while
the true model contains relevant covariates.
For the sake of convenience, we use generic notation $S\subseteq S_{F}$ to represent an arbitrary candidate model with $S=\{S_{\alpha},S_{\beta}\}$, and then denote
 $\alpha_{S}=(\alpha_k: k\in S_{\alpha})\in \mR^{|S_{\alpha}|}$ and $\beta_{S}=(\beta_l: l\in S_{\beta})\in \mR^{|S_{\beta}|}$.

Based on (\ref{equ:quasi-log-likelihood}), the Bayesian information criterion is
\begin{equation}
    \mbox{BIC}(S)=-2\ell_{c}(\hat{\alpha}_{S},\hat{\beta}_{S})+df\times \log n, \nonumber
\end{equation}
where $\ell_{c}(\hat{\alpha}_{S},\hat{\beta}_{S})$ represents the concentrated quasi-loglikelihood function for
any subset model $S$.
$\hat{\alpha}_{S}=(\hat{\alpha}_{k,S_{\alpha}}: k\in S_{\alpha})$ and $\hat{\beta}_{S}=(\hat{\beta}_{l,S_{\beta}}: l\in S_{\beta})$
 are the maximum likelihood estimates of
$\alpha_{S}$ and $\beta_{S}$, respectively, and $df=|S|$ is the size of $S$. Removing the irrelevant constants involved in the concentrated log-likelihood function, we have
\begin{equation}
    \mbox{BIC}(S)=n\log\hat{\sigma^{2}}(\hat{\eta}(\hat\alpha_{S},\hat\beta_{S}),\hat\alpha_{S},\hat\beta_S)-2\log|\det\{S(\hat\alpha_S,\hat\beta_S)\}|+|S|\times \log n. \nonumber
\end{equation}Then the optimal model selected by the Bayesian information criterion is $S_{\mbox{BIC}}=\mbox{argmin}_{S\subset S_{F}} \mbox{BIC}(S)$.
To obtain the theoretical property of BIC, we introduce an additional condition here.
\begin{itemize}
    \item [(C6)] For any underfitted model, which means $S\subset S_{F}$ but $S\not\supset S_T $ , assume that there exists a positive constant $c_{\min}>0$, such that
 $\mathop{\min}\limits_{S\not\supset S_\top }\big[\inf_{(\alpha_{S_T},\beta_{S_T})}E\{n^{-1}\ell_{c}(\alpha_{S_T},\beta_{S_T})\}-\inf_{(\alpha_{S},\beta_{S})}E\{n^{-1}\ell_{c}(\alpha_{S},\beta_{S})\}\big]>c_{\min}$.
\end{itemize}
\noindent Condition (C6) indicates that the mean of the concentrated quasi-loglikelihood function of any underfitted
model is inferior to that of the true model (i.e., no underfitted model can fit better than the true model). Similar conditions can be found in linear regression models; see, e.g.,
Assumption 3 in Shi and Tsai (2002) and Condition 2 in Wang et al. (2007).
We now present the following result.
\bet
Under Conditions (C1)-(C6), we have $P(S_{\mbox{BIC}}=S_T)\rightarrow1$ as $n\rightarrow\infty$.
\eet
\noindent
The above theorem implies that the Bayesian information criterion can determine the true model consistently as long as $n$ tends to infinity.
To obtain the solution of $S_{BIC}$, we apply the backward elimination method (see, e.g., Zhang and Wang 2011), which
can reduce the computational complexity from $O(2^{d_1+d_2})$ to $O\{(d_1+d_2)^2\}$. Thus, $S_{BIC}$ is computable when $d_1+d_2$ is not very large.

\csubsection{Hypothesis Testing}

After obtaining the parameter estimator of $\theta=(\eta^\top ,\alpha^\top ,\beta^\top ,\sigma^{2})^\top $,
we perform the following three tests to examine the homogeneity of $\lambda$ and $\gamma$.

Test I. We first test the homogeneity of $\lambda$ and $\gamma$ by considering the null and alternative  hypotheses:

\begin{center}
$H_{0}: \gamma_{1}=\gamma_{2}=\cdots=\gamma_{n}$ and $\lambda_{1}=\lambda_{2}=\cdots=\lambda_{n}$\\
 \mbox{~vs.~} $H_{1}: \gamma_{i_1}\neq \gamma_{i_2}$ or $\lambda_{j_1}\neq\lambda_{j_2}$\mbox{~for some~} $i_1\neq i_2$ and  $j_1\neq j_2.$ \nonumber
\end{center}
Assuming that $\gamma_{i}(\alpha)=F(Z_{i}^\top \alpha)$,
$\lambda_{j}(\beta)=F(V_{j}^\top \beta)$,
$F(\cdot)$ is strictly monotone,
no intercept is included in $\alpha$,
and $\mZ$ and $\mV_{-1}$ are of full rank as defined in Subsection 2.1,
the above hypotheses are equivalent to
\begin{center}
     $H_{0}$: $\alpha_{1}=\cdots=\alpha_{d_1}=0$ and $\beta_{2}=\cdots=\beta_{d_2}=0$~\\
      \mbox{~vs.~} $H_{1}: \alpha_i\not=0$ or $\beta_j\not =0$ for some $i\geq 1$ or $j>1$.
\end{center}

Recall that $\theta=(\eta^\top, \alpha^\top, \beta^\top ,\sigma^{2})^\top $
and $\beta_1$ is an intercept. To test the null hypothesis, we let $\theta_{1}=(\eta^\top, \beta_{1}, \sigma^{2})^\top $ and
$\theta_{2}=(\alpha_{1},\cdots,\alpha_{d_1}, \beta_{2},\cdots,\beta_{d_{2}})^\top $.
With a slight abuse of notation, we reset
$\theta=(\theta_{1}^\top, \theta_{2}^\top )^\top$.
The notation, functions and equations  used in the following theorem and propositions are based on this new setting.
Let $\hat{\theta}^{(c)}$ be the constrained QMLE under the null hypothesis.
Then, the resulting quasi-Lagrange-multiplier test statistic is
\[T_{s}=\frac{1}{n}\Big\{\frac{\partial\ell(\hat{\theta}^{(c)})}{\partial\theta}\Big\}^\top\mathcal{I}^{-1}_n(\hat{\theta}^{(c)})\Big\{\frac{\partial\ell(\hat{\theta}^{(c)})}{\partial\theta}\Big\},\]
where
$\frac{\partial\ell(\hat{\theta}^{(c)})}{\partial\theta}$
and $\mathcal{I}^{-1}_n(\hat{\theta}^{(c)})$ are, respectively, the score function and
the Fisher information matrix, both evaluated at $\hat{\theta}^{(c)}$.
Subsequently, the asymptotic distribution of $T_{s}$ is given below.

\bet
Under Conditions (C1)-(C5) and the null hypothesis of $H_{0}$, $T_{s}$ is asymptotically
 distributed as $\sum_{l=1}^{p+d_{1}+d_{2}+1} \lambda_{l}(\theta,\mu_{3},\mu_{4}) \chi^{2}_{l}(1)$
 as $n\to\infty$, where $\lambda_{l}(\theta,\mu_{3},\mu_{4})$ is the $l$-th largest eigenvalue of  $\mathcal{K}^{1/2}(\theta,\mu_{3},\mu_{4})\{\mathcal{I}^{-1}(\theta)-\mathcal{I}_1(\theta)\}\mathcal{K}^{1/2}(\theta,\mu_{3},\mu_{4})$,
$\mathcal{K}^{1/2}(\theta,\mu_{3},\mu_{4})$ and  $\mathcal{I}_1(\theta)$ are defined in Section 2 of the supplementary material,
 and $\chi^{2}_{l}(1)$ are independent chi-square random variables with 1 degree of freedom for $l=1,\cdots,(p+d_1+d_2+1)$.
 Furthermore, $T_{s}$ is asymptotically $\chi^{2}(d_1+d_2-1)$ when $\varepsilon$ is normally distributed.
\eet
\noindent
If the null hypothesis is not rejected, then we can consider the classical spatial autoregressive model. Otherwise,
there exists heterogeneity
either among inward or outward influence indices, or both, which motivates us to conduct the next two tests.

Test II. To test the homogeneity of $\gamma$, we consider the null and alternative hypotheses:
\beq H_{0,\gamma}: \gamma_{1}=\gamma_{2}=\cdots=\gamma_{n} \mbox{~vs.~} H_{1,\gamma}: \gamma_{i_1}\neq \gamma_{i_2} \mbox{~for some~} i_1\neq i_2. \nonumber\eeq
Assuming that $\gamma_{i}(\alpha)=F(Z_{i}^\top \alpha)$, $F(\cdot)$ is strictly monotone, no intercept is included in $\alpha$, and $\mZ$
 is of full rank defined in Subsection 2.1, the above hypotheses are equivalent to
\beq H_{0,\alpha}: \alpha_{1}=\alpha_{2}=\cdots=\alpha_{d_1}=0 \mbox{~vs.~} H_{1,\alpha}: \alpha_i\not=0 \mbox{~for some~} i. \eeq

Under the null hypothesis of $H_{0,\alpha}$, we can obtain the constrained QMLE $\hat{\theta}_\alpha^{(c)}$,
whose associated quasi-log likelihood function is $\ell(\hat{\theta}_\alpha^{(c)})$. Accordingly, the quasi-Lagrange Multiplier test
statistic of $\alpha$ is
\[T_{s\alpha}=\frac{1}{n}\Big\{\frac{\partial\ell(\hat{\theta}_{\alpha}^{(c)})}{\partial\theta}\Big\}^\top\mathcal{I}^{-1}_n(\hat{\theta}_{\alpha}^{(c)})
\Big\{\frac{\partial\ell(\hat{\theta}_{\alpha}^{(c)})}{\partial\theta}\Big\},\]
and its theoretical property is given below.

\bep
Under Conditions (C1)-(C5) and the null hypothesis of
$H_{0,\alpha}$, $T_{s\alpha}$ is
asymptotically distributed as $\sum_{l=1}^{p+d_{1}+d_2+1} \lambda_{\alpha l}(\theta,\mu_{3},\mu_{4})$ $ \chi^{2}_{l}(1)$ as $n\to\infty$,
where $\lambda_{\alpha l}(\theta,\mu_{3},\mu_{4})$ is the $l$-th largest eigenvalue of
$\mathcal{K}^{1/2}(\theta,\mu_{3},\mu_{4})$ $\{\mathcal{I}^{-1}(\theta)-\mathcal{I}_{\alpha}(\theta)\}\mathcal{K}^{1/2}(\theta,\mu_{3},\mu_{4})$, $\mathcal{I}_{\alpha}(\theta)$ is defined in Section 2 of the supplementary material,
and $\chi^{2}_{l}(1)$ are independent chi-square random variables with 1 degree of freedom for $l=1,\cdots,(p+d_{1}+d_2+1)$.
Furthermore, under the  normality assumption of $\varepsilon$, $T_{s\alpha}$ is asymptotically $\chi^{2}(d_1)$.
\eep
\noindent
If the null hypothesis is rejected, then there exists heterogeneity among inward influence indices. Otherwise, we can consider the network
influence model proposed by Zou et al. (2021).

Test III. To test the homogeneity of $\lambda$, we consider the null and alternative  hypotheses:
\beq H_{0,\lambda}: \lambda_{1}=\lambda_{2}=\cdots=\lambda_{n} \mbox{~vs.~}  H_{1,\lambda}: \lambda_{j_1}\neq\lambda_{j_2} \mbox{~for some~} j_1\neq j_2. \nonumber\eeq
Assuming that $\lambda_{j}(\beta)=F(V_{j}^\top \beta)$, $F(\cdot)$ is strictly monotone, and $\mV_{-1}$ is of full rank, the above hypotheses are equal to
\beq H_{0,\beta}: \beta_{2}=\cdots=\beta_{d_2}=0 \mbox{~vs.~} H_{1,\beta}: \beta_j\not=0 \mbox{~for some~} j>1.\nonumber\eeq

Under the null hypothesis of $H_{0,\beta}$, we can obtain the constrained QMLE $\hat{\theta}_{\beta}^{(c)}$ and its associated quasi-loglikelihood function $\ell(\hat{\theta}_{\beta}^{(c)})$.
Accordingly, the quasi-Lagrange Multiplier test statistic of $\beta$ is
\[T_{s\beta}=\frac{1}{n}\Big\{\frac{\partial\ell(\hat{\theta}_{\beta}^{(c)})}{\partial\theta}\Big\}^\top\mathcal{I}^{-1}_n(\hat{\theta}_{\beta}^{(c)})\Big\{\frac{\partial\ell(\hat{\theta}_{\beta}^{(c)})}{\partial\theta}\Big\}, \]
and its theoretical property is given below.

\bep
Under Conditions (C1)-(C5) and the null hypothesis of $H_{0,\beta}$, $T_{s\beta}$ is asymptotically distributed as $\sum_{l=1}^{p+d_{1}+d_{2}+1} \lambda_{\beta l}(\theta,\mu_{3},\mu_{4}) \chi^{2}_{l}(1)$ as $n\to\infty$, where $\lambda_{\beta l}(\theta,\mu_{3},\mu_{4})$ is the $l$-th largest eigenvalue of $\mathcal{K}^{1/2}(\theta,\mu_{3},\mu_{4})\{\mathcal{I}^{-1}(\theta)-\mathcal{I}_{\beta}(\theta)\}\mathcal{K}^{1/2}(\theta,\mu_{3},\mu_{4})$, $\mathcal{I}_{\beta}(\theta)$ is defined in Section 2
of the supplementary material, and $\chi^{2}_{l}(1)$ are independent chi-square random variables with 1 degree of freedom for $l=1,\cdots,(p+d_{1}+d_2+1)$. Furthermore, $T_{s\beta}$ is asymptotically $\chi^{2}(d_2-1)$ under the normality assumption of $\varepsilon$.
\eep
\noindent
If the null hypothesis is rejected, then there exists heterogeneity among outward influence indices.

It is worth noting that the asymptotic results in Theorem 3 and Propositions 1 and 2
depend on unknown parameters.
In practice, we can replace them by their corresponding consistent estimators.
For example, in the test statistic $T_{s}$,
$\lambda_{l}(\theta,\mu_{3},\mu_{4})$ is unknown. We can replace it by its consistent estimator
$\lambda_{n,l}(\hat{\theta}^{(c)},\hat{\mu}_{(3,c)},\hat{\mu}_{(4,c)})$, which is the $l$-th largest eigenvalue of $\mathcal{K}_{n}^{1/2}(\hat{\theta}^{(c)},\hat{\mu}_{(3,c)},\hat{\mu}_{(4,c)})\{\mathcal{I}_{n}^{-1}(\hat{\theta}^{(c)})
-\mathcal{I}_{n1}(\hat{\theta}^{(c)})\}\mathcal{K}_{n}^{1/2}(\hat{\theta}^{(c)},\hat{\mu}_{(3,c)},\hat{\mu}_{(4,c)})$.
Here, $\mathcal{K}_{n}(\hat{\theta}^{(c)},\hat{\mu}_{(3,c)},\hat{\mu}_{(4,c)})=\mathcal{I}_{n}(\hat{\theta}^{(c)})+\mathcal{J}_{n}(\hat{\theta}^{(c)},\hat{\mu}_{(3,c)},\hat{\mu}_{(4,c)})$ is a consistent estimator of $\mathcal{K}(\theta,\mu_{3},\mu_{4})$,
where  $\hat{\mu}_{(s,c)}=n^{-1}\sum_{i=1}^{n}(\hat{\varepsilon_i}^{(c)})^{s}$ for $s=3,4$ and  $\hat{\varepsilon}^{(c)}_{i}$ is the $i$-th element of
$\varepsilon(\hat{\eta}^{(c)},\hat{\alpha}^{(c)},\hat{\beta}^{(c)})$.
In addition, $\mathcal{I}_{n1}(\hat{\theta}^{(c)})$ is  a consistent
estimator of $\mathcal{I}_{1}(\theta)$.
Analogous replacements can be done for the asymptotic distributions of
$T_{s\alpha}$ and $T_{s\beta}$, which are omitted here.

\csection{NUMERICAL STUDIES}

\csubsection{Simulation Studies}

In this section, we conduct simulation studies to investigate the finite sample performance of parameter estimators, variable selection, and test statistics.
Let the off-diagonal elements, $a_{ij}$, of the adjacency matrix $A$  be independent and identically generated from the Bernoulli distribution with probability $5/n$,
and let the diagonal elements, $a_{ii}$, of $A$ be zeros. We then define the weighted adjacency matrix $W$ as $W=(w_{ij})_{n\times n}\in \mR^{n\times n}$ with $w_{ij}=a_{ij}/\sum_{j=1}^{n}a_{ij}$ for $i,j=1,\cdots,n$.
Next, the elements of $Z_{i}=(z_{i1},z_{i2}, z_{i3})^\top$, $V_{i}=(v_{i1},v_{i2}, v_{i3},v_{i4})^\top$ and $X_{i}=(x_{i1},x_{i2}, x_{i3})^\top$  are independent and
identically generated from the standard normal distribution $N(0, 1)$
except $v_{i1}=x_{i1}=1$.
Their corresponding coefficients are
$\alpha=(\alpha_{1},\alpha_{2}, \alpha_{3})^\top=(0.4, 0.6, 0.7)^\top$, $\beta=(\beta_{1},\beta_{2},\beta_{3},\beta_{4})^\top=(1.2, 1.4,1.3, 1.6)^\top$, and
 $\eta=(\eta_{1},\eta_{2},\eta_{3})^\top=(1,2,3)^\top$.
Subsequently, we have $\Gamma=
\mbox{diag}\{F(Z_{1}^\top\alpha),\cdots, F(Z_{n}^\top\alpha)\}$  and $\Lambda=\mbox{diag}\{F(V_{1}^\top\beta),\cdots, F(V_{n}^\top\beta)\}$.
Because $Z_{i}$ does not include the intercept, the above covariate-parameter setting ensures the model is identifiable; see Subsection 2.1.
In addition, the random errors $\varepsilon_i$ ($i=1,\cdots,n$) are independent and identically generated from two distributions:
the standard normal distribution, $N(0,1)$,  and the mixture normal
distribution, $0.9N(0,5/9)+0.1N(0,5)$, respectively. Finally,
the response vector $\mathbb{Y}$ is generated from model (2.2) with
$\mY=(I_n-\Gamma W\Lambda)^{-1}(\mX \eta+\epsilon)$ and parameter vector $\theta=(\eta^\top,\alpha^\top,\beta^\top,\sigma^2)^\top$.
All of these settings satisfy Conditions (C1)-(C6).

In simulation studies,
all  results are based on
500 realizations with $n$=500, 1,000 and 2,000.
Let $\hat{\theta}^{(k)}=(\hat{\eta}_{1}^{(k)},\hat{\eta}_{2}^{(k)},\hat{\eta}_{3}^{(k)},\hat{\alpha}_{1}^{(k)},
\hat{\alpha}_{2}^{(k)},\hat{\alpha}_{3}^{(k)},
\hat{\beta}_{1}^{(k)},\hat{\beta}_{2}^{(k)},\hat{\beta}_{3}^{(k)},\hat{\beta}_{4}^{(k)},
\hat{\sigma}^{2(k)})^\top\in\mathbb{R}^{11}$ be the QMLE  of
$\theta$ at the $k$-th realization and  let $\hat{\theta}_{j}^{(k)}$ be its $j$-th component.
To evaluate the performance of the
parameter estimates, we consider the measurement
$\mbox{BIAS}(\hat{\theta}_{j}^{(k)})=500^{-1}\sum_{k}(\hat{\theta}_{j}^{(k)}-\theta_{j})$, which is the averaged bias of $\hat{\theta}_{j}^{(k)}$.
Based on $\hat\theta^{(k)}$, we can further obtain the estimates $\hat\gamma^{(k)}$ and $\hat\lambda^{(k)}$ of $\gamma$ and $\lambda$, respectively. Their corresponding bias measurements are
$\mbox{BIAS}(\hat{\gamma})=n^{-1}\sum_{i=1}^n\{500^{-1}\sum_{k}(\hat{\gamma}_{i}^{(k)}-\gamma_{i})\}$ and $\mbox{BIAS}(\hat{\lambda})=n^{-1}\sum_{i=1}^n\{500^{-1}\sum_{k}(\hat{\lambda}_{i}^{(k)}-\lambda_{i})\}$.
By  Theorem 1, for each component $\hat{\theta}_{j}^{(k)}$, we can obtain the estimated standard error $\mbox{SD}_j^{(k)}$.
Thus, the average of the estimated standard errors for each $\theta_j$ is $\mbox{SD}_j=500^{-1}\sum_{k}\mbox{SD}_j^{(k)}$.
Subsequently, the averages of the estimated standard errors of the $n$ nodes via 500 realizations for $\gamma$ and $\lambda$ are
$\mbox{SD}_{\gamma}=n^{-1}\sum_{i=1}^n \big(500^{-1}\sum_{k}\mbox{SD}_{\gamma_i}^{(k)}\big)$ and $\mbox{SD}_{\lambda}=n^{-1}\sum_{i=1}^n \big(500^{-1}\sum_{k}\mbox{SD}_{\lambda_i}^{(k)}\big)$,  respectively.
To obtain an overall measurement,  we also consider the root mean squared error, $\mbox{RMSE}=\sqrt{\mbox{SD}^{2}+\mbox{BIAS}^{2}}$,  for $\theta$, $\gamma$ and $\lambda$.

Table 1 presents the BIAS, SD and RMSE of $\hat{\theta}$ ,$\hat{\gamma}$ and $\hat{\lambda}$ via 500 realizations under normal random errors with three sample sizes.
The results indicate that, for all three link functions, the absolute value of BIAS and the SD of the parameter estimates decrease
when $n$ gets large. In addition, the biases of $\gamma$ and $\lambda$ approach 0, and the average estimated errors are quite small. Hence, it is not surprising that RMSE reveals the same pattern. Notably, the accurate estimates of $\gamma$ and $\lambda$ allow us to subsequently conduct four-quadrant clustering, which is not reported here to save space.
Moreover, we find that both the estimated Hessian
matrix and the information matrix are positive definite at each iterative step across all realizations.  Accordingly, $\hat\theta$ is the global maximizer of the log-likelihood function, and
$\hat\theta$ is close to its true value, as shown in Table 1.
Under mixture normal errors,
the results  are qualitatively similar to those in Table 1, and we relegate them to Table S1  in the supplementary material.
As suggested by an anonymous reviewer, we  consider the simulation setting with an overlapping design. The results are qualitatively similar to those in Table 1, and we present them in Table S3 of the supplementary material.

We next study the finite sample performance of the BIC variable selection criterion. To this end,
we set up the full model size $|S_{F\alpha}|=|S_{F\beta}|$=6 and the true model size $|S_{T\alpha}|=3$, $|S_{T\beta}|$=4.
In addition,
three measures are used to assess the selection performance given below:
(i) The average percentage of correct fit (CF), i.e., $I(\hat{S}_{\alpha} =S_{T\alpha})$ and
$I(\hat{S}_{\beta} =S_{T\beta})$; (ii) the average true positive rate (TPR), i.e., $|\hat{S}_{\alpha}\cap S_{T\alpha}|/|S_{T\alpha}|$ and
$|\hat{S}_{\beta}\cap S_{T\beta}|/|S_{T\beta}|$; and (iii) the average false positive rate (FPR), i.e., $|\hat{S}_{\alpha}\cap S^{c}_{T\alpha}|/|S^{c}_{T\alpha}|$ and
$|\hat{S}_{\beta}\cap S^{c}_{T\beta}|/|S^{c}_{T\beta}|$, where $I$ is an indicator function, $\hat{S}_{\alpha}$ and $\hat{S}_{\beta}$ are the selected models via BIC,
 $S^{c}_{T\alpha}=S_{F\alpha}\backslash S_{T\alpha}$ and
$S^{c}_{T\beta}=S_{F\beta}\backslash S_{T\beta}$.

Table 2 reports CF, TPR and FPR under normal random errors via 500 realizations.
The results indicate that, as the sample size increases, the average percentage of CF and the average TPR  approach 1, while the average FPR is close to 0. Accordingly,  the optimal model selected by the BIC is consistently approaching the true model, which supports Theorem 2.
For the mixture normal errors,
the results are qualitatively similar to those in Table 2, and we present
them in  Table S2  of the supplementary material.

We finally investigate the performance of the three quasi Lagrange-multiplier tests by testing the homogeneity of the inward and outward influence indices
under normal random errors.
To this end, we set up the three hypotheses: (i) $\alpha=(0.4\delta,0.6\delta,0.7\delta)^\top$
and $\beta=(1.2,1.4\delta,1.3\delta,1.6\delta)^\top$, (ii) $\alpha=(0.4\delta,0.6\delta,0.7\delta)^\top$ and $\beta=(1.2,1.4,1.3,1.6)^\top$, and (iii) $\alpha=(0.4,0.6,0.7)^\top$ and
$\beta=(1.2,1.4\delta,1.3\delta,1.6\delta)^\top$, for Tests I, II, and III respectively with $\delta=0, 0.1, 0.15, 0.2, 0.25, 0.3, 0.4, 0.7$.
Here, $\delta=0$ evaluates the empirical sizes of the three tests, while other values of $\delta$ are used to assess the empirical powers.
Then, we calculate the empirical sizes and powers of $T_s$, $T_{s\alpha}$ and $T_{s\beta}$ under the three link functions with the significance level of 0.05 via 500 realizations.
Figures 2-4  demonstrate that the empirical sizes of the three tests approach the predetermined significance level, 0.05, as the sample size $n$ gets large, which demonstrates the validity of the three testing procedures mentioned in Subsection 2.4. In addition, the powers of the three tests increase and tend to 100\% when the sample size or $\delta$ becomes large. In sum, our proposed three tests perform well for assessing the heterogeneity of influence indices.

\csubsection{Real Data Analysis}

To illustrate the practical usefulness of the Inward and Outward Network Influence model and four-quadrant clustering method,
we collect social network data
from Sina Weibo, which is the biggest social media platform in China.
Our Sina Weibo dataset is a snapshot taken
on November 13, 2013 and contains
2,580 online users' activity and their corresponding attribute variables between January 1, 2013 and November 13, 2013, where all
users in our dataset had registered before January 1, 2013.
Accordingly, we define the adjacency matrix $A=(a_{ij})\in \mR^{2,580\times2,580}$ as follows: $a_{ij}=1$ if user $i$ is following user $j$
and $a_{ij}=0$ otherwise;
a similarly defined adjacency matrix can be found in Bramoull\'e et al. (2009), Zhou et al. (2017) and Zhu et al. (2019).
Since the network density is approximately equal to $1.3\%$, this network is extremely sparse.
Thus, Conditions (C2) and (C3) are satisfied.
The response variable $Y_{i}$ is the $i$-th user's activity level, which is proxied by the number of Sina Weibo posts made by user $i$.
However, on any given day,  we find that most users have an activity level (i.e., number of posts) of zero.
Thus our response variable is
the total number of posts over the entire study period.
In this way, our proposed model allows us
to assess the
overall inward and outward network influence
across the network during this period.


To characterize users' inward and outward influence, we consider the following five
attributes. (\romannumeral1) In-degree:
the number of each user's followers in the network; (\romannumeral2) Out-degree: the number of each user's followees in the network; (\romannumeral3) Duration:
the time elapsed since registering on the Sina Weibo platform; (\romannumeral4)
Self-Introduction: the length of a user's description about himself/herself;
(\romannumeral5) Gender: male=1 and female=0. We then assign all five attributes to
each of the three covariates $Z$, $V$ and $X$.
Afterwards, we fit them with the model (2.2),
and  find that
none of their associated coefficients $\alpha$, $\beta$ and $\eta$ are significant at the 5\% significance level.
Thus, we iteratively eliminated the attribute with  the largest $p$-value across $Z$, $V$ and $X$  until  one of the attributes becomes significant.
As a result,
the remaining attributes in
 $Z$, $V$ and $X$ are (``Out-degree", ``Duration" and ``Self-Introduction"),
(``In-degree", ``Duration" and ``Self-Introduction"), and
(``Gender"), respectively.

In this empirical example, the sample size is large and the full model is finite.
This motivates us to further employ the BIC consistent criterion to select the best model for each of the three link functions:
sigmoid, inverse of log-log and inverse of probit.
Inspired by Vuong (1989), we further choose the link function with the smallest BIC value, which is the inverse of probit.
Table 3 reports the QMLEs of the parameters and their corresponding $t$-statistics and $p$-values.
For $Z$ covariates, the attribute ``Out-degree'' is selected, and it is significant at the 5\% level with a positive coefficient.
This finding is sensible since users with more out-degree may have broad interests, outgoing personality, and strong curiosity to
accept new things and are thus more likely to be influenced by others.
As for $V$ covariates, the attribute ``Duration'' is selected, and it is significant at the 5\% level with a positive coefficient.
This result is also reasonable because the users with longer duration are generally called ``veteran users'', who tend to have greater influence in the network.
Finally,  the $p$-value of the $X$ variable ``gender'' is 0.1857. Thus, the users' activity level is not strongly related to gender.

It is worth noting that,
without employing an initial variable elimination procedure, none of the selected variables are significant at the 5\% significance level when using BIC to
select relevant variables from the full model with fifteen candidate variables (five attributes in each of three covariates $Z$, $V$ and $X$); the results are omitted to save space.
In addition, we find that both the estimated Hessian matrix and information matrix are positive definite at each iterative step, which ensures that the QMLEs listed in Table 3 are global maximizers.

We next examine the heterogeneity of inward and outward influence among individuals
via the three test statistics in Subsection 2.5. The $p$-values of all three tests are close to 0, which indicate high heterogeneity in both inward and outward influence
among the 2,580 users.
To evaluate these phenomena, we
calculate the estimated  influence indices $\hat{\gamma}_{i}=\Phi(Z_{i}^\top\hat{\alpha})$ and $\hat{\lambda}_{j}=\Phi(V_{j}^\top\hat{\beta})$.
We then sort them and obtain $\hat{\gamma}_{(1)}>\hat{\gamma}_{(2)}>\cdots>\hat{\gamma}_{(n)}$ and  $\hat{\lambda}_{(1)}>\hat{\lambda}_{(2)}>\cdots>\hat{\lambda}_{(n)}$.
 This allows us to conduct user segmentation. To this end,  as shown in Figure 1, we designate the origin as
 $\gamma_0=\hat\gamma_{(k_1)}$ and $\lambda_0=\hat\lambda_{(k_2)}$, where $k_1$ and $k_2$ satisfy
$\sum_{i=1}^{k_1}\hat{\gamma}_{(i)}/\sum_{i}\hat{\gamma}_{(i)}=12.5\%$ and
$\sum_{i=1}^{k_2}\hat{\lambda}_{(i)}/\sum_{i}\hat{\lambda}_{(i)}=12.5\%$, respectively.
To illustrate the usefulness of our proposed clustering method, we choose
the 12.5\% quantile as the threshold for our user segmentation.
In practice, decision makers can balance the benefits and costs of choosing a different threshold.

With the given  threshold values $\gamma_0$ and $\lambda_0$,
we now employ the four quadrant method of classifying
users, following two different approaches. First, we conduct the  ``indices-comparison (IC)'' by comparing
$\hat{\gamma}_{i}$ and $\hat{\lambda}_{i}$ with the corresponding  threshold
values $\gamma_0$ and $\lambda_0$  for $i=1,\cdots,n$.
Second, we proceed with the ``indices-test (IT)'' by
calculating the test statistics $T_{i,\gamma}$ and $T_{i,\lambda}$ for $i=1,\cdots,n$,
and obtain the corresponding $p$-values $p_{i,\gamma}$ and $p_{i,\lambda}$
by controlling the false discovery rate (FDR) as mentioned in Subsection 2.3.
Figure 5 depicts the percentage of users being classified into the four quadrants by the aforementioned two approaches.
Because the results are similar, we focus only on the ``indices-test''
approach.

Based on user segmentation, practitioners can design marketing strategies to improve a product promotion or opinion dissemination.
For example, if practitioners need to promote a new product on Weibo, they could focus on
not just ``influencers" but specifically
the 4.0\% ``early adopters". In contrast, if practitioners want to
conduct a promotion for a mature product, they should focus on the 7.7\% ``early majority".
To effectively utilize the above two types of users, decision makers can adopt the ``fan economy" strategy to increase their conversion rate.
With this strategy, they should first  focus on the 15.2\% ``opinion leaders" to make such promotions.
Finally,  removing or ignoring the 73.1\%  who are prolonged ``inactive users'' can reduce promoting costs and improve marketing efficiency.
In sum, this example demonstrates that obtaining the nodes' inward and outward influence indices can play an important role in social network applications.

\csection{CONCLUDING REMARKS}

In large scale networks, we propose an Inward and Outward Network Influence (IONI)
model that not only captures the heterogeneity among
nodes, but is also applicable to user segmentation.
Without imposing any specific error distribution, we apply the quasi-maximum likelihood approach to obtain the estimators of the inward and outward influence indices. This allows us to employ
the four-quadrant clustering method to segment users.
The theoretical properties of parameter estimates are established, and
simulation studies as well as an empirical analysis are presented to demonstrate the usefulness of the IONI model.

To broaden the usefulness of IONI and four-quadrant clustering,
we identify the following four possible future research avenues.
First, the adjacency matrix of the network is observed in our study.
For an unobserved adjacency matrix, we could allow it to follow a probability distribution and estimate the adjacency matrix. Then we could employ IONI for studying influential power within networks.
 It is not surprising that indirectly connected nodes can influence each other in a network; see Lan et al. (2018). Hence, the second avenue is to take into account higher-order adjacency matrices to enlarge the application of the IONI model.
Third, motivated by an anonymous reviewer's comment, one can consider  observations and network structures that vary with time. Hence, it is
useful to extend IONI to  dynamic network models; see, e.g., Dou et al. (2016) and Gao et al. (2019).
Lastly, based on an anonymous reviewer's suggestion, it is of great interest to generalize IONI to accommodate data with discrete responses (see Zhang et al. 2020).
We believe each of the above efforts would increase the value of IONI considerably.




\scsection{REFERENCES}

\begin{description}
\newcommand{\enquote}[1]{``#1''}
\expandafter\ifx\csname natexlab\endcsname\relax\def\natexlab#1{#1}\fi


\bibitem[{Anselin(1988)}]{Anselin:1988}
Anselin, L. (1988). \textit{Spatial Econometrics: Methods and Models}, Dordrecht: Kluwer Academic Press.

\bibitem[{Bamakan et al.(2019)}]{Bamakan:2019} Bamakan, S.~M.~H., Nurgaliev, I. and Qu, Q. (2019). \enquote{Opinion leader detection: A methodological review,}
 \textit{Expert Systems with Applications}, 115, 200--222.

\bibitem[{Banerjee(2013)}]{Banerjee:Chandrasekhar:Duflo:Jackson:2013}
Banerjee, A., Chandrasekhar, A. G., Duflo, E. and Jackson, M. O. (2013). \enquote{The diffusion of microfinance,}\textit{Science}, 341(6144).

\bibitem[{Banerjee(2014)}]{Banerjee:Chandrasekhar:Duflo:Jackson:2014}
Banerjee, A., Chandrasekhar, A. G., Duflo, E. and Jackson, M. O. (2014). \enquote{Gossip: Identifying central individuals in a social network,}\textit{National Bureau of Economic Research}, No. w20422.


\bibitem[{Bramoull\'e(2009)}]{Bramoule:2009}
Bramoull\'e, Y., Djebbari, H. and Fortin, B. (2009).
\enquote{Identification of peer effects through social networks,}
 \textit{Journal of Econometrics}, 150, 41--55.

\bibitem[{Carrington et al.(2005)}]{Carrington:2005}
Carrington, P. J., Scott, J. and Wasserman, S. (2005).
\textit{Models and Methods in Social Network Analysis}, Cambridge University Press.

\bibitem[{Catalini and Tucker(2017)}]{Catalini:2017} Catalini, C. and Tucker, C. (2017). \enquote{When early adopters don't adopt,}
 \textit{Science}, 357, 135--136.

\bibitem[{Dou et al.(2016)}]{Dou:Parrella:Yao:2016} Dou, B., Parrella, M.~L. and Yao, Q. (2016).
\enquote{Generalized Yule--Walker estimation for spatio-temporal models with unknown diagonal coefficients,} \textit{Journal of Econometrics}, 194, 369--382.

\bibitem[{Fan and Li(2001)}]{Fan:Li:2001}Fan, J. and Li, R. (2001).
\enquote{Variable selection via nonconcave penalized likelihood and its oracle properties,}
\textit{Journal of the American Statistical Association}, 96, 1348--1360.

\bibitem[{Gao et al.(2019)}]{Gao:2019} Gao, Z., Ma, Y., Wang, H. and Yao, Q. (2019).
\enquote{Banded spatio-temporal autoregressions,}
 \textit{Journal of Econometrics}, 208, 211-230.

\bibitem[{Gupta and Robinson(2015)}]{Gupta:Robinson:2015} Gupta, A. and Robinson, P.~M. (2015).
\enquote{Inference on higher-order spatial autoregressive models with increasingly many parameters,}
 \textit{Journal of Econometrics}, 186, 19--31.

\bibitem[{Hastie et al.(2009)}]{Hastie:2009}
Hastie, T., Tibshirani, R. and Friedman, J. (2009).
\textit{The Elements of Statistical Learning: Data Mining, Inference, and Prediction, Second Edition},  New York: Springer.

\bibitem[{Katona et al.(2011)}]{Katona:2011}
Katona, Z., Zubcsek, P.~P. and Sarvary, M. (2011). \enquote{Network effects and personal influences:
The diffusion of an online social network,}
\textit{Journal of Marketing Research}, 48, 425--443.

\bibitem[{Kumar and Reinartz(2018)}]{Kumar:2018}
Kumar, V. and Reinartz, V. (2018).
\textit{Customer Relationship Management: Concept, Strategy, and Tools}, Springer.

 \bibitem[{Kwok(2019)}]{Kwok:2019}
Kwok, H. H. (2019).
\enquote{Identification and estimation of linear social interaction models,}
 \textit{Journal of Econometrics}, 210, 434--458.

\bibitem[{Lam and Souza(2019)}]{Lam:Souza:2019} Lam, C. and Souza, P. (2019).
\enquote{Estimation and selection of spatial weight matrix in a spatial lag model,}
  \textit{Journal of Business $\&$ Economic Statistics}, 1--41.

\bibitem[{Lan et al.(2018)}]{Lan:Fang:Wang:Tsai:2018} Lan, W., Fang, Z., Wang, H. and Tsai, C.-L. (2018).
\enquote{Covariance matrix estimation via network structure,}
  \textit{Journal of Business $\&$ Economic Statistics}, 36, 359--369.

\bibitem[{Lawyer(2015)}]{Lawyer:2015} Lawyer, G. (2015).
\enquote{Understanding the influence of all nodes
in a network,}
  \textit{Nature: Scientific Reports}, 5: 8665.

\bibitem[{Lee(2004)}]{Lee:2004}
Lee, L. F. (2004). \enquote{Asymptotic distributions of quasi-maximum likelihood estimators for
spatial autoregressive models,} \textit{Econometrica}, 72, 1899--1925.

\bibitem[{LeSage and Pace(2009)}]{LeSage:Pace:2009}
LeSage, J. and Pace, R.~K. (2009). \textit{Introduction to Spatial
  Econometrics}, New York: Chapman \& Hall.

\bibitem[{Li et al.(2013)}]{Li:Ma:Zhang:Huang:2013} Li, Y., Ma, S., Zhang, Y. and Huang, R. (2013).
\enquote{An improved mix framework for opinion leader identification in online learning communities,}  \textit{Knowledge-Based Systems}, 43, 43--51.

\bibitem[{Liu and Lee(2010)}]{Lee:2010}
Liu, X. D. and Lee, L. F. (2010).
\enquote{GMM estimation of social interaction models with centrality,}
 \textit{Journal of Econometrics}, 159, 99--115.

\bibitem[{Ma and Liu(2014)}]{Ma:Liu:2014} Ma, N. and Liu, Y. (2014).
\enquote{SuperedgeRank algorithm and its application in identifying opinion leader of online public opinion supernetwork,}
 \textit{Expert Systems with Applications}, 41, 1357--1368.

\bibitem[{Maggio et al.(2019)}]{Maggio:2019} Maggio, M.~D., Franzoni, F., Kermani, A. and Sommavilla, C. (2019).
\enquote{The relevance of broker networks for information diffusion in the stock market,} \textit{Journal of Financial Economics}, 134, 419--446.

\bibitem[{Manski(1993a)}]{Manski:1993a}
Manski, C. F. (1993a).
\enquote{Identification of endogenous social effects: The reflection problem,}
 \textit{The Review of Economic Studies}, 60, 531--542.

  \bibitem[{Manski(1993b)}]{Manski:1993b}
Manski, C. F. (1993b).
\enquote{Identification problems in the social sciences,}
 \textit{American Sociological Association}, 23, 1--56.

\bibitem[{Mattila et al.(2003)}]{Mattila:Karjaluoto:Pento:2013} Mattila, M., Karjaluoto, H. and Pento, T. (2003).
 \enquote{Internet banking adoption among mature customers: early majority or laggards?}
 \textit{Journal of Services Marketing}, 17, 514--528.

\bibitem[{Ram and Jung(1994)}]{Ram:Jung:2012} Ram, S. and Jung, H.~S. (1994). \enquote{Innovativeness in product usage: A comparison of early adopters and early majority,}
\textit{Psychology $\&$ Marketing}, 11, 57--67.

\bibitem[{Scott(2013)}]{Scott:2013}
Scott, J. (2013). \textit{Social Network Analysis, Third Edition,} London: SAGE.

\bibitem[{Shi and Tsai(2002)}]{Shi:Tsai:2002} Shi, P. and Tsai, C.-L. (2002). \enquote{Regression model selection--A residual likelihood approach,}
 \textit{Journal of the Royal Statistical Society, Series B}, 64, 237--252.

\bibitem[{Storey et al.(2004)}]{Storey:Taylor:Siegmund:2004} Storey, J.~D., Taylor, J.~E. and Siegmund, D. (2004).
\enquote{Strong control, conservative point estimation and simultaneous conservative consistency of false discovery rates: a unified approach,}
\textit{Journal of the Royal Statistical Society: Series B}, 66, 187--205.

\bibitem[{Tabassum et al.(2018)}]{Tabassum:2018} Tabassum, S., Pereira, F.~S.~F., Fernandes, S. and Gama, J. (2018). \enquote{Social network analysis: An overview,}
 \textit{Wiley Interdisciplinary Reviews: Data Mining and Knowledge Discovery}, 8, e1256.

\bibitem[{Trusov et al.(2010)}]{Trusov:2010} Trusov, M., Bodapati, A. and Bucklin, R. (2010). \enquote{Determining influential users in Internet social networks,}
 \textit{Journal of Marketing Research}, 47, 643-658.


\bibitem[{Valente(2010)}]{Valente:2010}
Valente, T.~W. (2010). \textit{Social Networks and Health: Models, Methods, and Applications 1st Edition}, Oxford University Press.

\bibitem[{Vuong(1989)}]{Vuong:1989} Vuong, Q. H. (1989). \enquote{Likelihood ratio tests for model selection and non-nested hypotheses,}
 \textit{Econometrica}, 57, 307--333.

\bibitem[{Wang et al.(2007)}]{Wang:Li:Tsai:2007} Wang, H., Li, R. and Tsai, C.-L. (2007). \enquote{Tuning parameter selectors for the smoothly clipped absolute deviation method,}
\textit{Biometrika}, 94, 553--568.

\bibitem[{Wolf and Seebauer(2014)}]{Wolf:Seebauer:2014} Wolf, A. and Seebauer, S. (2014). \enquote{Technology adoption of electric bicycles: A survey among early adopters,}
 \textit{Transportation Research Part A: Policy and Practice}, 69, 196--211.

\bibitem[{Yu et al.(2008)}]{Yu:2008} Yu, J., Robert, de Jong. and Lee, L. (2008). \enquote{Quasi-maximum
 likelihood estimators for spatial dynamic panel data with fixed effects when both $n$ and $T$ are large,}
 \textit{Journal of Econometrics}, 146, 118--134.

\bibitem[{Zhang and Wang(2011)}]{Zhang:2011}
Zhang, Q. and Wang, H. (2011). \enquote{On BIC's selection consistency for discriminant analysis,}
\textit{Statistica Sinica}, 21(2): 731--740.

\bibitem[{Zhang and Yu(2018)}]{Zhang:Yu:2018}
  Zhang, X. and Yu, J. (2018).
\enquote{Spatial weights matrix selection and model averaging for spatial autoregressive models,}
 \textit{Journal of Econometrics}, 203, 1--18.

\bibitem[{Zhang et al.(2020)}]{Zhang:2020} Zhang, X., Pan, R., Guan, G., Zhu, X. and Wang, H. (2020). \enquote{Logistic regression with network data,}
 \textit{Statistica Sinica}, 30, 673--693.

\bibitem[{Zhou et al.(2017)}]{Zhou:Tu:Chen:Wang:2017} Zhou, J., Tu, Y., Chen, Y. and Wang, H. (2017). \enquote{Estimating spatial autocorrelation with sampled network data,}
\textit{Journal of Business $\&$ Economics Statistics}, 35, 130--138.

 \bibitem[{Zhu et al.(2017)}]{Zhu2017}
Zhu, X., Pan, R., Li, G., Liu, Y. and Wang, H. (2017). \enquote{Network vector autoregression,}
\textit{The Annals of Statistics}, 45, 1096--1123.

 \bibitem[{Zhu et al.(2019)}]{Zhu:2019}
Zhu, X., Chang, X., Li, R. and Wang, H. (2019). \enquote{Portal nodes screening for large scale social networks,}
\textit{Journal of Econometrics}, 209, 145--157.

\bibitem[{Zhu et al.(2020)}]{Zhu2020}
Zhu, X., Huang, D., Pan, R. and Wang, H. (2019). \enquote{Multivariate spatial autoregressive model for large scale social networks,}
\textit{Journal of Econometrics}, 215, 591--606.

\bibitem[{Zou et al.(2021)}]{Zou:Luo:Lan:Tsai:2021}
Zou, T., Luo, R., Lan, W. and Tsai, C.-L. (2021). \enquote{Network influence analysis,}
\textit{Statistica Sinica}, In Press.

\end{description}

\newpage
\makeatletter\def\@captype{table}\makeatother

\begin{spacing}{1}
\scriptsize
\centering
\setlength{\belowcaptionskip}{0.2cm}

\begin{landscape}
\caption{ The BIAS, SD and RMSE of the QMLEs ($\eta_{1}=1$, $\eta_{2}=2$, $\eta_{3}=3$, $\alpha_{1} =0.4$, $\alpha_{2} =0.6$, $\alpha_{3} = 0.7$, $\beta_{1}=1.2$, $\beta_{2}=1.4$, $\beta_{3}=1.3$, $\beta_{4}=1.6$ and $\sigma^{2}=1$)
under the three link functions. The random errors are simulated from the normal distribution $N(0,\sigma^{2})$.}
\label{Tab:01}
\begin{tabular}{c|ccccccccccccccc}
\hline
Link & $n$ & Measure & $\hat{\eta}_{1}$ & $\hat{\eta}_{2}$ & $\hat{\eta}_{3}$ & $\hat{\alpha}_{1}$ & $\hat{\alpha}_{2}$ & $\hat{\alpha}_{3}$ & $\hat{\gamma}$ & $\hat{\beta}_{1}$ & $\hat{\beta}_{2}$ & $\hat{\beta}_{3}$ & $\hat{\beta}_{4}$ &$\hat{\lambda}$ & $\hat{\sigma}^{2}$\\
\hline
\multirow{12}{*}{Sigmoid}&
\multirow{4}{*}{500}\\
& & BIAS &0.001 &0.001 &0.001 &0.002 &0.006 &-0.002 &0.000 &-0.008 &0.006 &-0.007 &-0.008 &-0.001 &-0.006\\
 & & SD &0.045 &0.045 &0.045 &0.128 &0.137 &0.142 &0.002 &0.391 &0.452 &0.428 &0.487 &0.007 &0.063\\
 & & RMSE &0.045 &0.045 &0.045 &0.128 &0.137 &0.142 &0.002 &0.391 &0.452 &0.428 &0.487 &0.007 &0.063\\
 \cline{2-16}
 & \multirow{4}{*}{1000}\\
& & BIAS &0.000 &0.002 &0.000 &-0.002 &-0.002 &0.001 &0.000 &-0.002 &-0.001 &0.005 &0.000 &-0.001 &-0.004\\
 & & SD &0.032 &0.032 &0.032 &0.089 &0.096 &0.100 &0.001 &0.273 &0.313 &0.301 &0.341 &0.003 &0.045\\
 & & RMSE &0.032 &0.032 &0.032 &0.089 &0.096 &0.100 &0.001 &0.273 &0.313 &0.301 &0.341 &0.003 &0.045\\
 \cline{2-16}
& \multirow{4}{*}{2000}\\
& & BIAS &0.000 &-0.001 &0.000 &-0.001 &-0.001 &-0.001 &0.000 &0.000 &-0.001 &0.007 &-0.003 &0.000 &-0.002\\
& & SD &0.022 &0.022 &0.022 &0.063 &0.068 &0.070 &0.000 &0.192 &0.220 &0.211 &0.239 &0.002 &0.032\\
& & RMSE &0.022 &0.022 &0.022 &0.063 &0.068 &0.070 &0.000 &0.192 &0.220 &0.211 &0.239 &0.002 &0.032\\
 \hline
 \multirow{12}{*}{Inverse of log-log}&
\multirow{4}{*}{500}\\
& & BIAS &0.001 &0.001 &0.001 &0.000 &0.004 &-0.002 &0.000 &0.000 &0.007 &-0.001 &-0.003 &0.000 &-0.006\\
 & & SD &0.045 &0.045 &0.045 &0.081 &0.091 &0.097 &0.001 &0.380 &0.392 &0.366 &0.433 &0.004 &0.063\\
 & & RMSE &0.045 &0.045 &0.045 &0.081 &0.091 &0.097 &0.001&	0.380 &0.392 &0.366 &0.433 &0.004 &0.063\\
 \cline{2-16}
 & \multirow{4}{*}{1000}\\
& & BIAS &0.000 &0.002 &0.000 &-0.001 &-0.003 &0.002 &0.000 &0.000 &0.002 &-0.002 &0.000 &0.000 &-0.004\\
 & & SD &0.032 	&0.032 &0.032 &0.056 &0.064 &0.069 &0.001 &0.263 &0.271 &0.255 &0.299 &0.002 &0.045\\
 & & RMSE &0.032 &0.032 &0.032 &0.056 &0.064 &0.069 &0.001 &0.263 &0.271 &0.255 &0.299 &0.002 &0.045\\
 \cline{2-16}
& \multirow{4}{*}{2000}\\
& & BIAS &0.000 &-0.001 &0.000 &-0.001&	-0.001 &0.000 &0.000 &0.000 &-0.001 &0.003&	0.001&	0.000 &-0.002\\
& & SD &0.023 &0.022 &0.022 &0.040 &0.045 &0.048 &0.000 &0.184 &0.189 &0.179 &0.211 &0.001 &0.032\\
& & RMSE &0.023 &0.022 &0.022 &0.040 &0.045 &0.048 &0.000 &0.184 &0.189 &0.179 &0.211 &0.001 &0.032\\
 \hline
 \multirow{12}{*}{Inverse of probit}&
\multirow{4}{*}{500}\\
& & BIAS &0.001 &0.001 &0.001 &0.001 &0.005 &-0.002 &0.000 &-0.006 &0.009 &-0.006 &-0.010 &-0.001 &-0.006\\
 & & SD &0.045 &0.045 &0.045 &0.088 &0.100 &0.107 &0.002 &0.377 &0.448 &0.416 &0.492 &0.006 &0.063\\
 & & RMSE &0.045 &0.045 &0.045 &0.088 &0.100 &0.107 &0.002&	0.377 &0.448 &0.416 &0.492&	0.007 &0.063\\
 \cline{2-16}
 & \multirow{4}{*}{1000}\\
& & BIAS &0.000 &0.002 &0.000 &0.000 &-0.002 &0.001 &0.000 &-0.001 &-0.001&	0.007 &-0.002 &0.000 &-0.004\\
 & & SD &0.032&	0.032 &0.032 &0.062 &0.070 &0.076 &0.001 &0.262 &0.309 &0.292 &0.342 &0.003 &0.045\\
 & & RMSE &0.032 &0.032 &0.032 &0.062 &0.070 &0.076 &0.001 &0.262 &0.309 &0.292 &0.342 &0.003 &0.045\\
 \cline{2-16}
& \multirow{4}{*}{2000}\\
& & BIAS &0.000 &-0.001 &0.000 &-0.001 &-0.001 &0.000 &0.000 &0.000 &-0.001 &0.004 &0.000 &0.000 &-0.002\\
& & SD &0.022 &0.022 &0.022 &0.043 &0.050 &0.053 &0.000 &0.184 &0.216 &0.204 &0.240 &0.001&	0.032\\
& & RMSE &0.022 &0.022 &0.022 &0.043 &0.050 &0.053 &0.000 &0.184 &0.216 &0.204 &0.240 &0.002 &0.032\\
 \hline
\end{tabular}
\end{landscape}
\end{spacing}

\newpage
\begin{spacing}{1}
\setlength{\belowcaptionskip}{0.2cm}
\scriptsize
\centering
\caption{ Variable selection results including (\%) of CF, TPR and FPR with  normal random errors $N(0,1)$.}

\label{Tab:02}
\begin{tabular}{lllll|lll}
\toprule
\multirow{2}{*}{Link function} & \multirow{2}{*}{$n$} &\multicolumn{3}{l}{$\hat{S_{\alpha}}$} & \multicolumn{3}{l}{$\hat{S_{\beta}}$}\\
\cline{3-8}
~\\
& & CF & TPR & FPR & CF & TPR & FPR\\
\midrule
\multirow{3}{*}{Sigmoid}
& 500 & 73.7 &90.4 &0.4 &79.3 &94.3 &0.2\\
 & 1000 &94.7&98.6 &0.3 &98.3 &99.7 &0.2\\
 & 2000 &100.0 &100.0 &0.0 &100.0 &100.0 &0.0\\
 \hline
 \multirow{3}{*}{Inverse of log-log}
& 500 &97.7 &99.3 &0.4 &98.7 &99.8 &0.3\\
& 1000 &99.0 &100.0 &0.3 &99.7 &100.0 &0.2\\
& 2000 &99.7 &100.0 &0.1 &100.0 &100.0 &0.0\\
 \hline
\multirow{3}{*}{Inverse of probit}
& 500 &95.3 &98.6 &0.1 &98.0 &99.6 &0.2\\
 & 1000 &99.7 &100.0 &0.1 &100.0 &100.0 &0.0\\
 & 2000 &100.0 &100.0 &0.0 &100.0 &100.0 &0.0\\
 \bottomrule
\end{tabular}
\end{spacing}

\setlength{\abovecaptionskip}{1cm}
\setlength{\belowcaptionskip}{0.2cm}
\begin{spacing}{1}
\scriptsize
\centering
\caption{ Variable selection results from the Sina Weibo data with the inverse of the probit link function.}
\label{Tab:04}
\begin{tabular}{c|ccccc}
\hline
 attribute-types & variables & estimates & standard error & t-statistic & p-value\\
\hline
\multirow{3}{*}{Z}\\
 & out-degree & 0.1263 & 0.0181 & 6.9747 & $\ll0.0001$ \\
 \cline{1-6}
\multirow{3}{*}{V}\\
 & duration & 0.6705 & 0.1356 & 4.9463 & $\ll0.0001$ \\
 \cline{1-6}
\multirow{4}{*}{X}\\
& intercept & 2.7460 & 0.1137 & 24.1494 & $<0.0001$ \\
 & gender & 0.0611& 0.0684 & 0.8939& 0.1857 \\
 \hline
\end{tabular}
\end{spacing}

~\\

\begin{figure}[H]
 \centering
 \includegraphics[width=5.6in,height=2in]{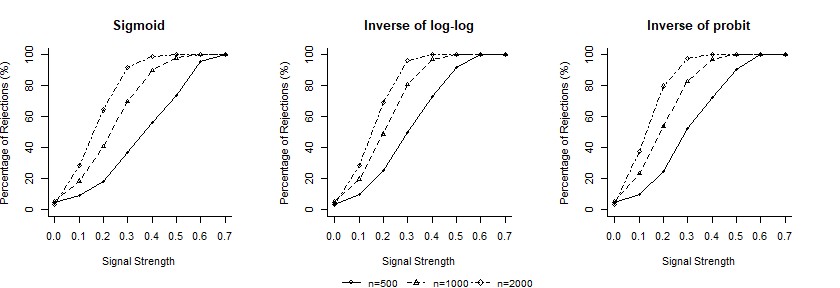}
\caption{The empirical sizes and powers of Test I under the three link functions with  significance level 0.05. The signal strengths are $\delta$=0, 0.1, 0.15, 0.2, 0.25, 0.3, 0.4 and 0.7, which correspond to the settings of $\alpha=(0.4\delta,0.6\delta,0.7\delta)^{\top}$ and $\beta = (1.2,1.4\delta,1.3\delta,1.6\delta)^{\top}$. The random errors are independently and identically simulated from the normal distribution $N(0,1)$.}
 \label{fig2}
\end{figure}

\begin{figure}[H]
 \centering
 \includegraphics[width=5.6in,height=2in]{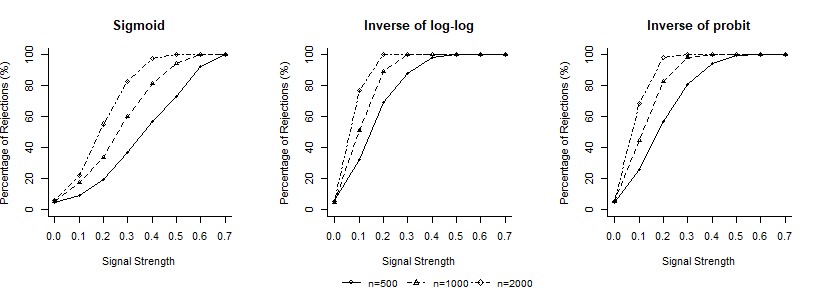}
 \caption{The empirical sizes and powers of Test II under the three link functions with  significance level 0.05. The signal strengths are $\delta$=0, 0.1, 0.15, 0.2, 0.25, 0.3, 0.4 and 0.7, which correspond to the settings of $\alpha=(0.4\delta,0.6\delta,0.7\delta)^{\top}$ and $\beta = (1.2, 1.4,  1.3, 1.6)^{\top}$. The random errors are independently and identically simulated from the normal distribution $N(0,1)$. }
 \label{fig3}
\end{figure}

\begin{figure}[H]
 \centering
 \includegraphics[width=5.6in,height=2in]{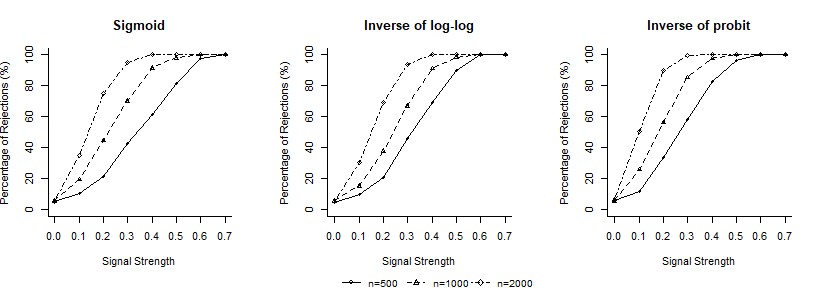}
 \caption{The empirical sizes and powers of Test III under the three link functions with  significance level 0.05. The signal strengths are $\delta$=0, 0.1, 0.15, 0.2, 0.25, 0.3, 0.4 and 0.7, which correspond to the settings of $\alpha=(0.4,0.6,0.7)^{\top}$ and $\beta = (1.2,1.4\delta,1.3\delta,1.6\delta)^{\top}$. The random errors are independently and identically simulated from the normal distribution $N(0,1)$.}
 \label{fig4}
\end{figure}

\begin{figure}[H]
 \centering
 \includegraphics[width=3.3in,height=2.3in]{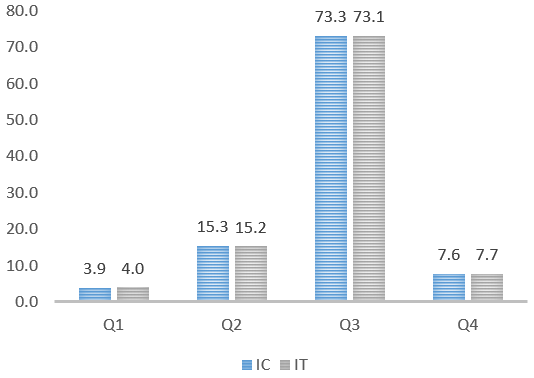}
 \caption{The proportion (\%) of each user segment in the Sina Weibo data as classified by the IC and IT methods. The Q1, Q2, Q3, and Q4
 represent the first quadrant (``early adopters"), the second quadrant (``opinion leaders"), the third quadrant (``inactive users"), and the fourth quadrant (``early majority"), respectively.  }
 \label{fig5}
\end{figure}

\end{document}


\begin{center}
{\bf\Large Inward and Outward Network Influence Analysis} \\
{Supplementary Material} \\
\bigskip
%
%

\end{center}

\renewcommand{\theequation}{S.\arabic{equation}}

\setcounter{equation}{0}

In this supplementary material, we include six sections.
Section 1 presents detailed expressions of the
Fisher information matrix $\mathcal{I}_n(\theta)$ and the quantity $\mathcal{J}_n(\theta,\mu_3,\mu_4)$
used in estimating the asymptotic covariance matrix of parameter estimators.  Additional notations used for proving Theorem 3 and Propositions 1-2 are given in Section 2.
Section 3 provides five technical lemmas.
Section 4 presents the proofs of theorems and propositions.
Simulation studies for mixture normal errors and  for an overlapping design are provided in Sections 5 and 6, respectively. Both additional studies are used
to demonstrate the robustness of our proposed estimators.


\scsubsection{Section 1: Detailed Expressions of $\mathcal{I}_n(\theta)$ and $\mathcal{J}_n(\theta,\mu_3,\mu_4)$}

Let $\Gamma_{\alpha_{k}}(\alpha):=\partial\Gamma(\alpha)/\partial\alpha_{k}=\mbox{diag}\big\{z_{1k}F'(Z_{1}^\top \alpha),\cdots,z_{nk}F'(Z_{n}^\top \alpha)\big\}$ for $k = 1,\cdots,d_{1}$,   and $\Lambda_{\beta_{l}}(\beta):=\partial\Lambda(\beta)/\partial\beta_{l}=\mbox{diag}\big\{v_{1l}F'(V_{1}^\top \alpha),\cdots,v_{nl}F'(V_{n}^\top \beta)\big\}$ for $l = 1,\cdots,d_2$. The Fisher information matrix $\mathcal{I}(\theta)$ of the quasi-maximum likelihood function $\ell(\theta)$ is denoted as follows:
\begin{equation}
\mathcal{I}_{n}(\theta):=-n^{-1}E\left[\frac{\partial^{2}\ell(\theta)}{\partial\theta\partial\theta^\top }\right]=\begin{pmatrix}
\sigma^{-2}n^{-1}\mathbb{X}^\top \mathbb{X} & \mathcal{I}_{\eta\alpha,n} & \mathcal{I}_{\eta\beta,n} &0_{p\times1}\\
\mathcal{I}_{\alpha\eta,n} & \mathcal{I}_{\alpha\alpha,n}  & \mathcal{I}_{\alpha\beta,n}  & \mathcal{I}_{\alpha\sigma^{2},n}\\
\mathcal{I}_{\beta\eta,n} & \mathcal{I}_{\beta\alpha,n}  & \mathcal{I}_{\beta\beta,n}  & \mathcal{I}_{\beta\sigma^{2},n}\\
0_{1\times p} & \mathcal{I}_{\sigma^{2}\alpha,n}  & \mathcal{I}_{\sigma^{2}\beta,n}  & 2^{-1}\sigma^{-4}\\
\end{pmatrix}\label{equ:Fisher-matrix},\nonumber
\end{equation}
where
\begin{center}
$\mathcal{I}_{\eta\alpha,n}=\frac{1}{n\sigma^{2}}\left(\mX^\top \Gamma_{\alpha_{1}}(\alpha) W\Lambda(\beta) S^{-1}(\alpha,\beta)\mX\eta,\cdots,\mX^\top \Gamma_{\alpha_{d_{1}}}(\alpha)W\Lambda(\beta) S^{-1}(\alpha,\beta)\mX\eta\right)_{p\times d_{1}}$,
~\\
$\mathcal{I}_{\eta\beta,n}=\frac{1}{n\sigma^{2}}\left(\mX^\top \Gamma(\alpha)W\Lambda_{\beta_{1}}(\beta)S^{-1}(\alpha,\beta)\mX\eta,\cdots,\mX^\top \Gamma(\alpha)W\Lambda_{\beta_{d_{2}}}(\beta)S^{-1}(\alpha,\beta)\mX\eta\right)_{p\times d_{2}}$,
~\\
$\mathcal{I}_{\alpha\alpha,n}=n^{-1}\Big(tr\big\{\Gamma_{\alpha_{k_{1}}}(\alpha) W\Lambda(\beta)
S^{-1}(\alpha,\beta)\Gamma_{\alpha_{k_{2}}}(\alpha) W\Lambda(\beta) S^{-1}(\alpha,\beta)\big\}$
$ +tr\big\{\Gamma_{\alpha_{k_{1}}}(\alpha) W\Lambda(\beta) S^{-1}(\alpha,\beta)S^{-1}(\alpha,\beta)^\top \Lambda(\beta) W^\top  \Gamma_{\alpha_{k_{2}}}(\alpha)\big\}$
$ +\sigma^{-2}{\eta^\top \mX^\top S^{-1}(\alpha,\beta)^\top \Lambda(\beta) W^\top \Gamma_{\alpha_{k{2}}}(\alpha)\Gamma_{\alpha_{k{1}}}(\alpha)W\Lambda(\beta) S^{-1}(\alpha,\beta)\mX\eta}\Big) _{d_{1}\times d_{1}}$,~\\
~\\
$\mathcal{I}_{\alpha\beta,n}=n^{-1}\Big(tr\big\{\Gamma_{\alpha_{k}}(\alpha) W\Lambda(\beta) S^{-1}(\alpha,\beta)\Gamma(\alpha) W\Lambda_{\beta_{l}}(\beta)S^{-1}(\alpha,\beta)\big\}$
$ +tr\big\{\Gamma_{\alpha_{k}}(\alpha) W\Lambda(\beta) S^{-1}(\alpha,\beta)S^{-1}(\alpha,\beta)^\top \Lambda_{\beta_{l}}(\beta) W^\top  \Gamma(\alpha)\big\}$
$ +\sigma^{-2}{\eta^\top \mX^\top S^{-1}(\alpha,\beta)^\top \Lambda_{\beta_{l}}(\beta) W^\top \Gamma(\alpha)\Gamma_{\alpha_{k}}(\alpha)W\Lambda(\beta) S^{-1}(\alpha,\beta)\mX\eta}\Big) _{d_{1}\times d_{2}}$,~\\
~\\
$\mathcal{I}_{\beta\beta,n}=n^{-1}\Big(tr\big\{\Gamma(\alpha) W\Lambda_{\beta_{l_{1}}}(\beta) S^{-1}(\alpha,\beta)\Gamma(\alpha) W\Lambda_{\beta_{l_{2}}}(\beta) S^{-1}(\alpha,\beta)\big\}$
$ +tr\big\{\Gamma(\alpha) W\Lambda_{\beta_{l_{1}}}(\beta) S^{-1}(\alpha,\beta)S^{-1}(\alpha,\beta)^\top \Lambda_{\beta_{l_{2}}}(\beta) W^\top  \Gamma(\alpha)\big\}$
$ +\sigma^{-2}{\eta^\top \mX^\top S^{-1}(\alpha,\beta)^\top \Lambda_{\beta_{l_{2}}}(\beta) W^\top \Gamma(\alpha)\Gamma(\alpha)W\Lambda_{\beta_{l_{1}}}(\beta) S^{-1}(\alpha,\beta)\mX\eta}\Big) _{d_{2}\times d_{2}}$,~\\
~\\
$\mathcal{I}_{\alpha\sigma^{2},n}=\sigma^{-2}n^{-1}\Big(tr\big\{S^{-1}\Gamma_{\alpha_{1}}(\alpha) W\Lambda(\beta)\},\cdots,tr\{S^{-1}\Gamma_{\alpha_{d_{1}}}(\alpha) W\Lambda(\beta)\big\}\Big)^\top  _{d_{1}\times1}$,~\\
~\\
$\mathcal{I}_{\beta\sigma^{2},n}=\sigma^{-2}n^{-1}\Big(tr\big\{S^{-1}\Gamma(\alpha)W\Lambda_{\beta_{1}}(\beta)\},\cdots,tr\{S^{-1}\Gamma(\alpha) W\Lambda_{\beta_{d_{2}}}(\beta)\big\}\Big)^\top  _{d_{2}\times1}$,~\\
~\\
$\mathcal{I}_{\alpha\eta,n}=\mathcal{I}_{\eta\alpha,n}^\top,$
$\mathcal{I}_{\beta\eta,n}=\mathcal{I}_{\eta\beta,n}^\top, $
$\mathcal{I}_{\beta\alpha,n}=\mathcal{I}_{\alpha\beta,n}^\top $, $\mathcal{I}_{\sigma^{2}\alpha,n}=\mathcal{I}_{\alpha\sigma^{2},n}^\top $, and $\mathcal{I}_{\sigma^{2}\beta,n}=\mathcal{I}_{\beta\sigma^{2},n}^\top.$
\end{center}

Let $\circ$ denote the Hadamard product of matrices, $l_{n}=(1,\cdots,1)^\top \in \mR^{n\times1}$,
$\mathbb{X}_{j} = (X_{1j},\cdots,X_{nj})^\top \in \mR^{n}$ for $j = 1,\cdots,p$. In addition, denote
the third and the fourth order moments by $\mu_{3} =
\textrm{E}(\epsilon_{i}^{3})$ and $\mu_{4} =\textrm{E}(\epsilon_{i}^{4})$, respectively, for $i = 1,\cdots,n$. After algebraic simplification, we have
\begin{center}
$\mathcal{J}_{n}(\theta,\mu_{3},\mu_{4})=\begin{pmatrix}
0_{p\times p} & \mathcal{J}_{\eta\alpha,n} & \mathcal{J}_{\eta\beta,n} & \frac{\mu_{3}\mathbb{X}^\top l_{n}}{2n\sigma^{6}}\\
\mathcal{J}_{\alpha\eta,n} & \mathcal{J}_{\alpha\alpha,n}  & \mathcal{J}_{\alpha\beta,n}  & \mathcal{J}_{\alpha\sigma^{2},n}\\
\mathcal{J}_{\beta\eta,n} & \mathcal{J}_{\beta\alpha,n}  & \mathcal{J}_{\beta\beta,n}  & \mathcal{J}_{\beta\sigma^{2},n}\\
\frac{\mu_{3}l_{n}^\top \mathbb{X}}{2n\sigma^{6}} & \mathcal{J}_{\sigma^{2}\alpha,n}  & \mathcal{J}_{\sigma^{2}\beta,n}  & \frac{\mu_{4}-2\sigma^{4}}{4\sigma^{8}}\\
\end{pmatrix}$, where~\\
~\\
$\mathcal{J}_{\eta\alpha,n}=\frac{\mu_{3}}{n\sigma^{4}}\Big(tr\big[\big(\mX_{j}l_{n}^\top \big)\circ\big\{\Gamma_{\alpha_{k}}\big(\alpha)W\Lambda(\beta)S^{-1}(\alpha,\beta)\big\}\big]\Big)_{p\times d_{1}}$,
~\\
~\\
$\mathcal{J}_{\eta\beta,n}=\frac{\mu_{3}}{n\sigma^{4}}\Big(tr\big[\big(\mX_{j}l_{n}^\top \big)\circ\big\{\Gamma(\alpha)W\Lambda_{\beta_{l}}(\beta)S^{-1}(\alpha,\beta)\big\}\big]\Big)_{p\times d_{2}}$,
~\\
~\\
$\mathcal{J}_{\alpha\alpha,n}=\frac{\mu_{4}-3\sigma^{4}}{n\sigma^{4}}\Big(tr\big[\big\{\Gamma_{\alpha_{k_{1}}}
(\alpha)W\Lambda(\beta)S^{-1}(\alpha,\beta)\big\}\circ\big\{\Gamma_{\alpha_{k_{2}}}(\alpha)W\Lambda(\beta)S^{-1}
(\alpha,\beta)\big\}\big]\Big)+\frac{\mu_{3}}{n\sigma^{4}}\Big(tr\big[\big\{\Gamma_{\alpha_{k_{1}}}(\alpha)W
\Lambda(\beta)S^{-1}\mX\eta l_{n}^\top \big\}\circ\big\{\Gamma_{\alpha_{k_{2}}}(\alpha)W\Lambda(\beta)S^{-1}(\alpha,\beta)\big\}\big]\Big)+\frac{\mu_{3}}{n\sigma^{4}}
\Big(tr\big[\big\{\Gamma_{\alpha_{k_{2}}}(\alpha)W\Lambda(\beta)S^{-1}(\alpha,\beta)\mX\eta l_{n}^\top \big\}\circ\big\{\Gamma_{\alpha_{k_{1}}}(\alpha)W\Lambda(\beta)S^{-1}(\alpha,\beta)\big\}\big]\Big)_{d_{1}\times d_{1}}$,~\\
~\\
$\mathcal{J}_{\alpha\beta,n}=\frac{\mu_{4}-3\sigma^{4}}{n\sigma^{4}}\Big(tr\big[\big\{\Gamma_{\alpha_{k}}(\alpha)W
\Lambda(\beta)S^{-1}(\alpha,\beta)\big\}\circ\big\{\Gamma(\alpha)W\Lambda_{\beta_{l}}(\beta)
S^{-1}(\alpha,\beta)\big\}\big]\Big)+\frac{\mu_{3}}{n\sigma^{4}}\Big(tr\big[\big\{\Gamma_{\alpha_{k}}(\alpha)W
\Lambda(\beta)S^{-1}\mX\eta l_{n}^\top \big\}\circ\big\{\Gamma(\alpha)W\Lambda_{\beta_{l}}(\beta)S^{-1}(\alpha,\beta)\big\}\big]\Big)+\frac{\mu_{3}}{n\sigma^{4}}
\Big(tr\big[\big\{\Gamma(\alpha)W\Lambda_{\beta_{l}}(\beta)S^{-1}(\alpha,\beta)\mX\eta l_{n}^\top \big\}\circ\big\{\Gamma_{\alpha_{k}}(\alpha)W\Lambda(\beta)S^{-1}(\alpha,\beta)\big\}\big]\Big)_{d_{1}\times d_{2}}$,~\\
~\\
$\mathcal{J}_{\beta\beta,n}=\frac{\mu_{4}-3\sigma^{4}}{n\sigma^{4}}\Big(tr\big[\big\{\Gamma(\alpha)W
\Lambda_{\beta_{l_{1}}}(\beta)S^{-1}(\alpha,\beta)\big\}\circ\big\{\Gamma(\alpha)W\Lambda_{\beta_{l_{2}}}
(\beta)S^{-1}(\alpha,\beta)\big\}\big]\Big)+\frac{\mu_{3}}{n\sigma^{4}}\Big(tr\big[\big\{\Gamma(\alpha)W
\Lambda_{\beta_{l_{1}}}(\beta)S^{-1}\mX\eta l_{n}^\top \big\}\circ\big\{\Gamma(\alpha)W\Lambda_{\beta_{l_{2}}}(\beta)S^{-1}(\alpha,\beta)\big\}\big]\Big)+\frac{\mu_{3}}{n\sigma^{4}}
\Big(tr\big[\big\{\Gamma(\alpha)W\Lambda_{\beta_{l_{2}}}(\beta)S^{-1}(\alpha,\beta)\mX\eta l_{n}^\top \big\}\circ\big\{\Gamma(\alpha)W\Lambda_{\beta_{l_{1}}}(\beta)S^{-1}(\alpha,\beta)\big\}\big]\Big)_{d_{2}\times d_{2}}$,
\end{center}
\[\mathcal{J}_{\alpha\sigma^{2},n}=\frac{\mu_{4}-3\sigma^{4}}{2n\sigma^{6}}
\Big(tr\big\{\Gamma_{\alpha_{k}}(\alpha)W\Lambda(\beta)S^{-1}(\alpha,\beta)\big\}\Big)+\frac{\mu_{3}}{2n\sigma^{6}}
\Big(l_{n}^\top \Gamma_{\alpha_{k}}(\alpha)W\Lambda(\beta)S^{-1}(\alpha,\beta)\mX\eta\Big)_{d_{1}\times1},\]
\[\mathcal{J}_{\beta\sigma^{2},n}=\frac{\mu_{4}-3\sigma^{4}}{2n\sigma^{6}}
\Big(tr\big\{\Gamma(\alpha)W\Lambda_{\beta_{l}}(\beta)S^{-1}(\alpha,\beta)\big\}\Big)+\frac{\mu_{3}}{2n\sigma^{6}}
\Big(l_{n}^\top \Gamma(\alpha)W\Lambda_{\beta_{l}}(\beta)S^{-1}(\alpha,\beta)\mX\eta\Big)_{d_{2}\times1},\]
$\mathcal{J}_{\alpha\eta,n}=\mathcal{J}_{\eta\alpha,n}^\top $,
$\mathcal{J}_{\beta\eta,n}=\mathcal{J}_{\eta\beta,n}^\top $,
$\mathcal{J}_{\beta\alpha,n}=\mathcal{J}_{\alpha\beta,n}^\top $, $\mathcal{J}_{\sigma^{2}\alpha,n}=\mathcal{J}_{\alpha\sigma^{2},n}^\top $, and $\mathcal{J}_{\sigma^{2}\beta,n}=\mathcal{J}_{\beta\sigma^{2},n}^\top $.

\scsubsection{Section 2: Additional Notation used for Theorem 3 and Propositions 1-2}

\noindent{$\mathbf{Notation~used~for~Theorem~3.}$}
Recall that $\theta=(\eta^\top, \alpha^\top, \beta^\top ,\sigma^{2})^\top $
and $\beta_1$ is an intercept. In consideration for hypothesis testing, we let $\theta_{1}=(\eta^\top, \beta_{1}, \sigma^{2})^\top $ and
$\theta_{2}=(\alpha_{1},\cdots,\alpha_{d_1}, \beta_{2},\cdots,\beta_{d_{2}})^\top $.
With a slight abuse of notation, we reset
$\theta=(\theta_{1}^\top, \theta_{2}^\top )^\top$. In addition, let
\begin{center}
$\Delta_{T}=\begin{pmatrix}I_{p}  & 0_{p\times 1}  & 0_{p\times1} &0_{p\times d_{1}} & 0_{p\times (d_{2}-1)}\\
0_{1\times p}   & 1 & 0 & 0_{1\times d_{1}} & 0_{1\times(d_{2}-1)} \\
0_{1\times p}   &0 & 1 & 0_{1\times d_{1}} & 0_{1\times (d_{2}-1)} \\\end{pmatrix}
\in \mR^{(p+2)\times(p+d_{1}+d_{2}+1)}$
\end{center}
and $\mathcal{I}_{11}(\theta)=\Delta_
{T} \mathcal{I}(\theta)\Delta^\top_{T}$. Then,
\begin{center}
  $\mathcal{I}_{11}^{-1}(\theta)=(\Delta_{T} \mathcal{I}(\theta)\Delta^\top_{T} )^{-1}=:
\begin{pmatrix}
l_{11}(\theta) & l_{12}(\theta)\\
l_{21}(\theta) & l_{22}(\theta)
\end{pmatrix},$
\end{center}
where $l_{11}(\theta) \in \mR^{(p+1)\times (p+1)}$, $l_{12}(\theta) \in \mR^{(p+1)\times1}$, $l_{21}(\theta) \in \mR^{1\times (p+1)}$ and $l_{22}(\theta) \in \mR^{1\times1}$.
Finally, define
\begin{equation}
\mathcal{I}_1(\theta)=\begin{pmatrix}
 l_{11}(\theta)  & l_{12}(\theta) & 0_{(p+1)\times (d_1+d_2-1)}\\
 l_{21}(\theta)  & l_{22}(\theta) & 0_{1\times (d_1+d_2-1)}\\
  0_{(d_1+d_2-1)\times (p+1)} & 0_{(d_1+d_2-1)\times 1} & 0_{(d_1+d_2-1)\times (d_1+d_2-1)}\nonumber
     \end{pmatrix}.
\end{equation}

\noindent{$\mathbf{Notation~for~Propositions~1.}$} Analogously, we reset
$\theta=(\theta_{1}^\top, \theta_{2}^\top )^\top$ with $\theta_{1}=(\eta^\top, \beta^\top, \sigma^{2})^\top $ and $\theta_{2}=(\alpha_{1},\cdots,\alpha_{d_1})^\top $
in consideration for hypotheses testing in Test II. In addition, let
\begin{center}
$\Delta_{T_\alpha}=\begin{pmatrix}I_{p} & 0_{p\times d_{2}} & 0_{p\times1} &0_{p\times d_{1}}\\
0_{d_2\times p} & I_{d_{2}} & 0_{d_2\times1} & 0_{d_2\times d_{1}} \\
0_{1\times p}  & 0_{1\times d_{2}} & 1& 0_{1\times d_{1}} \\\end{pmatrix}
\in \mR^{(p+d_2+1)\times(p+d_{1}+d_{2}+1)}$
\end{center}
and
$\mathcal{I}_{11}(\theta_\alpha)=\Delta_
{T_\alpha} \mathcal{I}(\theta)\Delta^\top_{T_\alpha} $. Then,
\begin{center}
  $\mathcal{I}_{11}^{-1}(\theta_\alpha)=(\Delta_{T_\alpha} \mathcal{I}(\theta)\Delta^\top_{T_\alpha} )^{-1}=:
\begin{pmatrix}
l_{\alpha_{11}}(\theta) & l_{\alpha_{12}}(\theta)\\
l_{\alpha_{21}}(\theta) & l_{\alpha_{22}}(\theta)
\end{pmatrix},$
\end{center}
where $l_{\alpha_{11}}(\theta) \in \mR^{(p+d_2)\times (p+d_2)}$, $l_{\alpha_{12}}(\theta) \in \mR^{(p+d_2)\times1}$, $l_{\alpha_{21}}(\theta) \in \mR^{1\times (p+d_2)}$ and $l_{\alpha_{22}}(\theta) \in \mR^{1\times1}$.
Finally, define
\begin{equation}
     \mathcal{I}_{\alpha}(\theta)=\begin{pmatrix}
     l_{\alpha_{11}}(\theta) & l_{\alpha_{12}}(\theta) & 0_{(p+d_2)\times d_1} \\
    l_{\alpha_{21}}(\theta) & l_{\alpha_{22}}(\theta) & 0_{1\times d_1}\\
    0_{d_{1}\times (p+d_2)} & 0_{d_1\times 1} & 0_{d_1\times d_1} \nonumber
     \end{pmatrix}.
\end{equation}

\noindent{$\mathbf{Notation~for~Propositions~2.}$}
Based on the hypothesis testing in  Test III, we reset $\theta=(\theta_1,\theta_2)$ with $\theta_{1}=(\eta^\top, \alpha^\top,\beta_{1}, \sigma^{2})^\top $ and $\theta_{2}=(\beta_{2},\cdots,\beta_{d_{2}})^\top $.  In addition, let
\begin{center}
$\Delta_{T_\beta}=\begin{pmatrix}I_{p+d_1}  & 0_{(p+d_1)\times 1} & 0_{(p+d_1)\times1} & 0_{(p+d_1)\times (d_{2}-1)}\\
0_{1\times (p+d_1)}  & 1 & 0 & 0_{1\times (d_{2}-1)}\\
0_{1\times (p+d_1)} & 0 & 1 & 0_{1\times (d_{2}-1)}\\\end{pmatrix}
\in \mR^{(p+d_1+2)\times(p+d_{1}+d_{2}+1)}$,
\end{center}
and
$\mathcal{I}_{11}(\theta_\beta)=\Delta_
{T_\beta} \mathcal{I}(\theta)\Delta^\top_{T_\beta} $. Then,
\begin{center}
  $\mathcal{I}_{11}^{-1}(\theta_\beta)=(\Delta_{T_\beta} \mathcal{I}(\theta)\Delta^\top_{T_\beta} )^{-1}=:
\begin{pmatrix}
l_{\beta_{11}}(\theta) & l_{\beta_{12}}(\theta)\\
l_{\beta_{21}}(\theta) & l_{\beta_{22}}(\theta)
\end{pmatrix},$
\end{center}
where $l_{\beta_{11}}(\theta) \in \mR^{(p+d_1+1)\times (p+d_1+1)}$, $l_{\beta_{12}}(\theta) \in \mR^{(p+d_1+1)\times1}$, $l_{\beta_{21}}(\theta) \in \mR^{1\times (p+d_1+1)}$, $l_{\beta_{22}}(\theta) \in \mR^{1\times1}$.
Finally, define
\begin{equation}
     \mathcal{I}_{\beta}(\theta)=\begin{pmatrix}
     l_{\beta_{11}}(\theta) & l_{\beta_{12}}(\theta) & 0_{(p+d_1+1)\times(d_{2}-1)} \\
     l_{\beta_{21}}(\theta)  & l_{\beta_{22}}(\theta) & 0_{1\times (d_{2}-1)}\\
     0_{(d_{2}-1)\times (p+d_1+1)} & 0_{(d_{2}-1)\times 1} & 0_{(d_{2}-1)\times (d_{2}-1)} \nonumber
     \end{pmatrix}.
\end{equation}
It is worth noting that we use $\mathcal{K}(\theta,\mu_{3},\mu_{4})=\mathcal{I}(\theta)+\mathcal{J}(\theta,\mu_{3},\mu_{4})$ in all three tests.

\scsubsection{Section 3: Five Useful Lemmas}

Lemmas 1--5 can be found in Zou et al. (2021), although the detailed proofs of Lemma 5
 need to be modified from those in Lemma 3 of Zou et al. (2021).
They can be obtained  from the authors upon request.

\bel
For any generic vector $x=(x_{1},\cdots,x_{q})^\top \in \mR^{q}$ and for $1\leq s_{1}\leq s_{2}$, we have that
\begin{center}
    $\|x\|_{s_{1}}\leq\|x\|_{s_{2}}q^{1/s_{1}-1/s_{2}}\,\,\mbox{~and~}\,\,\begin{vmatrix}\sum\limits_{j=1}^{q}x_{j}\end{vmatrix}^{s_{1}}\leq q^{s_{1}-1}\sum\limits_{j=1}^{q}|x_{j}|^{s_{1}}$.
\end{center}
\eel

\bel
For any generic vector $x=(x_{1},\cdots,x_{q})^\top \in \mR^{q}$ and generic matrix $G\in \mR^{m\times q}$ and $U\in \mR^{m\times m}$ and for any $s\ge1$,
we have that $\|UGx\|_{s}\leq m^{1/s}|G|_{\infty}\|U\|_{\infty}\|x\|_{1}$.
\eel

\bel
Define $\varepsilon=(\varepsilon_{1},\cdots,\varepsilon_{n})^\top $, where $\varepsilon_{1},\cdots,\varepsilon_{n}$
are independent and identically distributed random variables with mean 0 and finite variance $\sigma^{2}$. Let
\begin{center}
    $Q_{n}=\varepsilon^\top A \varepsilon+b^\top \varepsilon-\sigma^{2}tr(A)$,
\end{center}
where $A=(a_{ij})_{n\times n}\in \mR^{n\times n}$ and $b=(b_{1},\cdots,b_{n})^\top \in \mR^{n\times1}$. Suppose that:
\begin{itemize}
\item [(1)] for $i,j=1,\cdots,n$, $a_{ij}=a_{ji}$;
\item [(2)] $\mathop{\sup}\limits_{n\ge 1}\|A\|_{1}<\infty$;
\item [(3)] for some $\xi_{1}>0,\mathop{\sup}\limits_{n\ge1}n^{-1}\|b\|_{2+\xi_{1}}^{2+\xi_{1}}<\infty$;
\item [(4)] for some $\xi_{2}>0$, $E|\varepsilon_{i}|^{4+\xi_{2}}<\infty$.
\end{itemize}
Then, we have $E(Q_{n})=0$ and
\begin{center}
    $\sigma^{2}_{Q_{n}}:=var(Q_{n})=4\sigma^{4}\sum\limits_{i=1}^{n}\sum\limits_{j=1}^{i-1}a_{ij}^{2}+\sigma^{2}\sum\limits_{i=1}^{n}b_{i}^{2}+
    \sum\limits_{i=1}^{n}\big\{(\mu_{4}-\sigma^{4})a_{ii}^{2}+2\mu_{3}b_{i}a_{ii}\big\},$
\end{center}
where $\mu_{s}=E(\varepsilon_{i}^{s})$ for $s=3,4$. Moreover, suppose $n^{-1}\sigma^{2}_{Q_{n}}\ge c$ for some $c\ge 0$.
Then, we obtain
\begin{center}
    $\sigma^{-1}_{Q_{n}}Q_{n}\xrightarrow{d}N(0,1)$.
\end{center}
\eel

\bel
Let $\varepsilon=(\varepsilon_{1},\cdots,\varepsilon_{n})^\top $, where $\varepsilon_{1},\cdots,\varepsilon_{n}$ are independent and identically distributed with mean 0 and finite variance $\sigma^{2}$. Define
\begin{center}
    $\mathcal{Q}_{n}=\begin{pmatrix}
    \vec^\top (A_{1})\\
    \vdots\\
    \vec^\top (A_{L})\end{pmatrix}\vec(\varepsilon\varepsilon^\top -\sigma^{2}I_{n})+B^\top \varepsilon,$
\end{center}
where $A_{l}=(a_{ij}^{(l)})_{n\times n}\in \mR^{n\times n}$ for $l=1,\cdots,L$ with $L<\infty$ and $B=(b_{il})_{n\times L}\in \mR^{n\times L}$. Suppose the following assumptions are satisfied,
\begin{itemize}
\item [(1)] for all $i,j=1,\cdots,n$ and $l=1,\cdots,L$, $a_{ij}^{(l)}=a_{ji}^{(l)}$;
\item [(2)] $\mathop{\sup}\limits_{n\ge1}\|A_{l}\|_{1}<\infty$ for any $l=1,\cdots,L$;
\item [(3)] for some $\xi_{1}>0$, $\mathop{\sup}\limits_{n\ge1}n^{-1}\|\vec(B)\|_{2+\xi_{1}}^{2+\xi_{1}}<\infty$;
\item [(4)] for some $\xi_{2}>0$, $E|\varepsilon_{i}|^{4+\xi_{2}}<\infty$.
\end{itemize}
Then, we have $E(\mathcal{Q}_{n})=0$ and
\begin{center}
    $\cov(\mathcal{Q}_{n})=2\sigma^{2}\big\{tr(A_{l_{1}}A_{l_{2}})\big\}_{L\times L}+\sigma^{2}B^\top B+\{\mu_{4}-3\sigma^{4}\}\Psi^\top \Psi+\mu_{3}\{\Psi^\top B+B^\top \Psi\}$,
\end{center}
where $\Psi=(\psi_{1},\cdots,\psi_{L})\in \mR^{n\times L}$ with $\psi
_{l}=(a_{11}^{l},\cdots,a_{nn}^{l})^\top \in \mR^{n}$ for $l=1,\cdots,L$, and $\mu_{s}=E(\varepsilon_{i}^{s})$ for $s=3,4$. Moreover, assume that
$n^{-1}\cov(\mathcal{Q}_{n})\to \mathcal{D}$ holds for a positive definite matrix $\mathcal{D}\in \mR^{L\times L}$. Then we have
\begin{center}
    $n^{-1/2}\mathcal{Q}_{n}\xrightarrow{d}N(0,\mathcal{D})$.
\end{center}
\eel

\bel
Under Conditions (C1)-(C5), we have that, as $n\to\infty$,
\[
(i.)\quad n^{-1/2}\frac{\partial\ell(\theta)}{\partial\theta}\xrightarrow{d}N\Big(0,\mathcal{I}(\theta)+\mathcal{J}(\theta,\mu_{3},\mu_{4})\Big), \quad (ii.)
-n^{-1}\frac{\partial^{2}\ell(\theta)}{\partial\theta\partial\theta^\top }\xrightarrow{p}\mathcal{I}(\theta),
\]
\eel

\scsubsection{Section 4: Proofs of the Theorems and Propositions}

\noindent{$\mathbf{Proof~of~Theorem~1.}$}
To show this theorem, we take following three steps:
 (1) demonstrating that $\hat{\theta}$ is $n^{1/2}$-consistent,
as $n\to\infty$; (2) verifying the asymptotically normal property of $\hat{\theta}$; (3) showing that $\hat{\gamma}_i$ and $\hat{\lambda}_i$ are also asymptotically normal for each node $i$. Detailed procedures are given below.

Step 1. Applying the techniques of Fan and Li (2001), we  need to show that, for an arbitrarily small positive constant $\xi>0$, there exists a constant $M_{\xi}>0$ that  satisfies
\begin{equation}
    P\Big\{
    \mathop{\sup}\limits_{u\in \mR^{(p+d_{1}+d_{2}+1)}:||u||_{2}=M_{\xi}}\ell(\theta+n^{-1/2}u)<\ell(\theta)
    \Big\}\ge1-\xi,
\end{equation}
as $n\to\infty$. To prove this, employing the Taylor series expansion and using the results of Lemma 5, we obtain
\begin{center}
$\mathop{\sup}\limits_{u\in \mR^{(p+d_{1}+d_{2}+1)}, \|u\|_{2}=M_{\xi}}
\begin{Bmatrix}\ell(\theta+n^{-1/2}u)-\ell(\theta)\end{Bmatrix}$~\\
\end{center}
\begin{center}
$ =\mathop{\sup}\limits_{u\in \mR^{(p+d_{1}+d_{2}+1)}, \|u\|_{2}=M_{\xi}}
\begin{bmatrix}
n^{-1/2}\frac{\partial\ell(\theta)}{\partial\theta^\top }\cdot u-(2n)^{-1}u^\top \big\{-\frac{\partial^{2}\ell(\theta)}{\partial\theta\partial\theta^\top }\big\}u+o_{p}(1)\end{bmatrix}$
\end{center}
\begin{equation}
\leq M_{\xi}\cdot O_{p}(1)-2^{-1}\lambda_{\min}\{\mathcal{I}(\theta)\}\cdot M^{2}_{\xi}+o_{p}(1),
\end{equation}
where $\lambda_{\min}\{\cdot\}$ is the smallest eigenvalue of the matrix inside the braces. Note that $M_{\xi}\cdot O_{p}(1)-2^{-1}\lambda_{\min}\{\mathcal{I}(\theta)\}\cdot M^{2}_{\xi}$
in (S.2) is a quadratic function of $M_{\xi}$. This, in conjunction with Condition (C5),  implies that $-2^{-1}\lambda_{\min}\{\mathcal{I}(\theta)\}<0$.
Thus, as long as $M_{\xi}$ is large enough, we have
\begin{equation}
    P\Big[
    \mathop{\sup}\limits_{u\in \mR^{(p+d_{1}+d_{2}+1)}, \|u\|_{2}=M_{\xi}}\big\{\ell(\theta+n^{-1/2}u)-\ell(\theta)
    \big\}< 0
    \Big]\to 1,
\end{equation}
which demonstrates (S.1). The above equation indicates that there exists a local maximizer $\hat{\theta}$ satisfying $\|\hat{\theta}-\theta\|_{2}\leq n^{-1/2}M_{\xi}$ as $n$ is large enough. This, together with (S.1), leads to
\begin{center}
$P\begin{pmatrix}\|\hat{\theta}-\theta\|_{2}\leq M_{\xi}/\sqrt{n}\end{pmatrix}\ge P\Big\{\mathop{\sup}\limits_{u\in \mR^{(p+d_{1}+d_{2}+1)}, \|u\|_{2}=M_{\xi}}\ell(\theta+n^{-1/2}u)<\ell(\theta)\Big\}\ge1-\xi.$
\end{center}
Accordingly, we  obtain $\sqrt{n}\|\hat{\theta}-\theta\|_{2}=O_{p}(1)$, which completes the first part of the proof.

Step 2. Applying the Taylor series expansion to the score function and the result of Step 1, we have that $0=\frac{\partial\ell(\hat{\theta})}{\partial\theta}=\frac{\partial\ell(\theta)}{\partial\theta}+\frac{\partial^{2}\ell(\theta)}{\partial\theta\partial\theta^\top }(\hat{\theta}-\theta)+O_{p}(1)$, which implies
\[\hat{\theta}-\theta=-\Big\{\frac{\partial^{2}\ell(\theta)}{\partial\theta\partial\theta^\top}\Big\}^{-1}
    \Big\{\frac{\partial\ell(\theta)}{\partial\theta}+O_{p}(1)\Big\}=
    -\Big\{\frac{1}{n}\frac{\partial^{2}\ell(\theta)}{\partial\theta\partial\theta^\top }\Big\}^{-1}
    \Big\{\frac{1}{n}\frac{\partial\ell(\theta)}{\partial\theta}+o_{p}(1)\Big\}.\]
Combing this with the results of Lemma 5, we obtain
\begin{equation}
    \sqrt{n}(\hat{\theta}-\theta)=\frac{1}{\sqrt{n}}\mathcal{I}^{-1}(\theta)\frac{\partial\ell(\theta)}{\partial\theta}+
    o_{p}(1)\xrightarrow{d}N\begin{pmatrix}0,\mathcal{I}^{-1}(\theta)+\mathcal{I}^{-1}(\theta)\mathcal{J}(\theta)\mathcal{I}^{-1}(\theta)\end{pmatrix},
\end{equation}
which completes the second part of the proof.

Step 3.
Let $l_\alpha=\big(0_{d_1\times p},I_{d_1}, 0_{d_1\times (d_2+1)}\big)$ and $l_\beta=\big(0_{d_2\times (p+d_1)},I_{d_2}, 0_{d_2\times 1}\big)$. Denote
$\mathcal{D}_\alpha(\theta,\mu_{3},\mu_{4})=
 l_\alpha\big\{\mathcal{I}^{-1}(\theta)+\mathcal{I}^{-1}(\theta)\mathcal{J}(\theta,\mu_{3},\mu_{4})\mathcal{I}^{-1}(\theta)\big\}l_\alpha^\top$ and $\mathcal{D}_\beta(\theta,\mu_{3},\mu_{4})=
 l_\beta\big\{\mathcal{I}^{-1}(\theta)+\mathcal{I}^{-1}(\theta)\mathcal{J}(\theta,\mu_{3},\mu_{4})\mathcal{I}^{-1}(\theta)\bigl\} l_\beta^\top$.
Using the result of  (S.4), we have
\begin{center}
$\sqrt{n}(\hat{\alpha}-\alpha)\xrightarrow{d}N\begin{pmatrix}0,\mathcal{D}_\alpha(\theta,\mu_3,\mu_4)\end{pmatrix}$
and
$\sqrt{n}(\hat{\beta}-\beta)\xrightarrow{d}N\begin{pmatrix}0,\mathcal{D}_\beta(\theta,\mu_3,\mu_4)\end{pmatrix}.$
\end{center}
Since $\gamma_i= F(Z_i^\top\alpha)$ and $\lambda_i=F(V_i^\top\beta)$,
we  employ the delta method and obtain that
 \begin{equation}
\sqrt{n}(\hat{\gamma}_i-\gamma_i)\xrightarrow{d}N\begin{pmatrix}0,\big(F'(Z_i^\top\alpha)\big)^2Z_i^\top \mathcal{D}_\alpha(\theta,\mu_3,\mu_4)Z_i\end{pmatrix} ~~{\mbox{and}}\nonumber
 \end{equation}
 \begin{equation}
\sqrt{n}(\hat{\lambda}_i-\lambda_i)\xrightarrow{d}N\begin{pmatrix}0,\big(F'(V_i^\top\beta)\big)^2V_i^\top\mathcal{D}_\beta(\theta,\mu_3,\mu_4)V_i\end{pmatrix}, \nonumber
 \end{equation}
which completes the entire proof.

\noindent{$\mathbf{Proof~of~Theorem~2.}$} To prove this theorem, we consider the following two scenarios, i.e., the candidate model $S\subset S_{F}$  is overfitted or underfitted.

\textbf{Overfitted Model:}
$S\in\mathcal{Q}_{+}=\{S:S_T \subset S, S_T \ne S\}$.
Denote the QMLEs of $S_T$ and $S$ by $(\hat{\alpha}_{S_T}^\top,\hat{\beta}_{S_T }^\top)^\top\in \mR^{|S_T|}$ and $(\hat{\alpha}_{S}^\top,\hat{\beta}_{S}^\top)^\top\in \mR^{|S|}$,
respectively.
Without loss of generality, we
assume that the
$\alpha$ and $\beta$ components of $S_T$ correspond to the first $|S_T|$
components of $S$. Accordingly,  $(\tilde{\alpha}_{S}^\top,\tilde{\beta}_{S}^\top)=(\hat{\alpha}_{S_T}^\top,0^\top ,\hat{\beta}_{S_T}^\top,0^\top )^\top\in \mR^{|S|}$.
 By Conditions (C1)-(C5) and Theorem 1, we have that $(\tilde{\alpha}_{S}^\top,\tilde{\beta}_{S}^\top)^\top$ is $n^{1/2}$ consistent and $\ddot{\ell}_{c}(\hat{\alpha}_{S},\hat{\beta}_{S})$ is finite. Using the fact that
$\dot{\ell}_{c}(\hat{\alpha}_{S},\hat{\beta}_{S})=0$, we then  apply the Taylor series expansion and obtain
\[\ell_{c}(\hat{\alpha}_{S_T},\hat{\beta}_{S_T})-\ell_{c}(\hat{\alpha}_{S},\hat{\beta}_{S})
=\ell_{c}^*(\tilde{\alpha}_{S},\tilde{\beta}_{S})-\ell_{c}(\hat{\alpha}_{S},\hat{\beta}_{S})\]
\[=\big\{(\tilde{\alpha}_{S},
\tilde{\beta}_{S})-(\hat{\alpha}_{S},\hat{\beta}_{S})\big\}^\top
\ddot{\ell}_{c}(\hat{\alpha}_{S},\hat{\beta}_{S})\big\{(\tilde{\alpha}_{S},\tilde{\beta}_{S})
-(\hat{\alpha}_{S},\hat{\beta}_{S})\big\}\big\{1+o_{p}(1)\big\}\]
\[\ge\lambda_{\min}\big\{\ddot{\ell}_{c}(\hat{\alpha}_{S},\hat{\beta}_{S})\big\}\big\|(\hat{\alpha}_{S},\hat{\beta}_{S})
-(\tilde{\alpha}_{S},\tilde{\beta}_{S})\big\|_{2}^{2}\big\{1+o_{p}(1)\big\}=O_{p}(1),\]
where $\ell_{c}^*(\tilde{\alpha}_{S},\tilde{\beta}_{S})$ is the concentrated quasi-loglikelihood function of the overfitted model and it is evaluated at $(\tilde{\alpha}_{S},\tilde{\beta}_{S})$.
Thus, we obtain that
\[\mbox{BIC}(S)-\mbox{BIC}(S_T)=2\big\{\ell_{c}(\hat{\alpha}_{S_T},\hat{\beta}_{S_T})
-\ell_{c}(\hat{\alpha}_{S},\hat{\beta}_{S})\big\}+\big(|S|-|S_T|\big)\times\log n \]
\[\ge O_{p}(1)+\log n\to \infty.
\]
Since $\mathcal{Q}_{+}$ is a finite set with $|S_{F}|<\infty$, the above result implies
\[P\big[\inf_{S\in\mathcal{Q}_{+}}\big\{\mbox{BIC}(S)-\mbox{BIC}(S_\top)\big\}\ge 0\big]\to 1.
\]
Accordingly, for any $S\in\mathcal{Q}_{+}$, we have $P(S_{\mbox{BIC}}=S_T)\to 1$.

\textbf{Underfitted Model:} $S\in\mathcal{Q}_{-}=\{S:S_T \not\subset S\}$.
Using the result that $(\hat{\alpha}_{S_T}^\top,\hat{\beta}_{S_T}^\top)^\top$ is a consistent
estimator of $(\alpha_{S_T}^\top,\beta_{S_T}^\top)^\top$ as $n\to\infty$.
By the law of large numbers, we have that $n^{-1}\ell_{c}(\hat{\alpha}_{S_T},\hat{\beta}_{S_T})-E\big\{n^{-1}\ell_{c}(\alpha_{S_T},\beta_{S_T})\big\}=o_{p}(1)$.
Analogously, we can obtain that  $n^{-1}\ell_{c}(\hat{\alpha}_{S},\hat{\beta}_{S})-E\big\{n^{-1}\ell_{c}(\alpha_{S},\beta_{S})\big\}=o_{p}(1)$.
By Condition (C6), we have $n^{-1}\ell_{c}(\hat{\alpha}_{S_T},\hat{\beta}_{S_T})-n^{-1}\ell_{c}(\hat{\alpha}_{S},\hat{\beta}_{S})=E\big\{n^{-1}\ell_{c}(\alpha_{S_\top },\beta_{S_\top })-n^{-1}\ell_{c}(\alpha_{S},\beta_{S})\big\}+o_{p}(1)\ge c_{\min}+o_{p}(1)$. Accordingly, $n^{-1}\big\{\mbox{BIC}(S)-\mbox{BIC}(S_T)\big\}\ge c_{\min}+o_{p}(1)$, which implies
\[ P\big[\inf_{S\in\mathcal{Q}_{-}}\big\{\mbox{BIC}(S)-\mbox{BIC}(S_T)\big\}\ge 0\big]\to 1.\]
This, together with the overfitted result, leads to
\[P\big[\inf_{S\in\mathcal{Q}_{-}\cup\mathcal{Q}_{+}}\big\{\mbox{BIC}(S)\ge \mbox{BIC}(S_T)\big\}\big]=P\big(S_{\mbox{BIC}}=S_T\big)\to 1,\]
which completes the entire proof.

\noindent{$\mathbf{Proof~of~Theorem~3.}$}
As mentioned in Section 2, we set $\theta=(\theta_{1}^\top, \theta_{2}^\top )^\top$, where
$\theta_{1}=(\eta^\top, \beta_{1}, \sigma^{2})^\top $ and
$\theta_{2}=(\alpha_{1},\cdots,\alpha_{d_1}, \beta_{2},\cdots,\beta_{d_{2}})^\top $.
The following notation, functions, and equations are based on this new setting.
Under the null hypothesis, i.e., $H_0:\theta_{2} = (0,\cdots,0)^\top \in \mR^{(d_1+d_{2}-1)\times1}$,
the true parameter is $\theta=(\theta_{1}^\top, 0^\top )^\top$.
Denote the resulting constrained and unconstrained QMLE of $\theta$ as $\hat{\theta}^{(c)}$ and $\hat\theta$, respectively. That is, $\hat{\theta}^{(c)}$
is obtained by maximizing the resulting quasi-loglikelihood function with the
constraint that $\theta_2=0$, while $\hat\theta$ is obtained without such a constraint.
Furthermore, define
\begin{center}
    $\mathcal{I}(\theta)=\begin{pmatrix}
    \mathcal{I}_{11}(\theta) & \mathcal{I}_{12}(\theta)\\
    \mathcal{I}_{21}(\theta) & \mathcal{I}_{22}(\theta)
    \end{pmatrix},$
\end{center}
where $\mathcal{I}_{ij}(\theta)$ is the convergence of the corresponding information
matrix with respect to $\theta_{i}$ and $\theta_{j}$ for $i,j \in\{1,2\}$.

Under the null hypothesis  $H_0$, we employ similar techniques to those used in proving (S.4) to obtain that
\begin{equation}
    \sqrt{n}(\hat{\theta}^{(c)}-\theta)=\frac{1}{\sqrt{n}}\mathcal{I}_1(\theta)\frac{\partial\ell(\theta)}{\partial\theta}+o_{p}(1),
\nonumber
\end{equation}
where
 $\mathcal{I}_1(\theta)=\begin{pmatrix}
    \mathcal{I}^{-1}_{11}(\theta) & 0_{(p+d_1+2)\times(d_{2}-1)}\\
    0_{(d_{2}-1)\times(p+d_1+2)} & 0_{(d_{2}-1)\times(d_{2}-1)}
    \end{pmatrix}$
defined in Section 2.
Accordingly,  $\hat{\theta}^{(c)}$ is a $\sqrt n$-consistent estimator. This, together with (S.4), leads to
\begin{equation}
    \sqrt{n}(\hat{\theta}^{(c)}-\hat{\theta})=\frac{1}{\sqrt{n}}\big\{\mathcal{I}_1(\theta)
    -\mathcal{I}^{-1}(\theta)\big\}\frac{\partial\ell(\theta)}{\partial\theta}+o_{p}(1)=O_{p}(1).
\end{equation}
In addition, applying the Taylor series expansion, we obtain that
\begin{center}
$n^{-1/2}\frac{\partial\ell(\hat{\theta}^{(c)})}{\partial\theta}=n^{-1}\frac{\partial^{2}\ell
(\check{\theta})}{\partial\theta\partial\theta^\top}n^{1/2}(\hat{\theta}^{(c)}-\hat{\theta})$,
\end{center}
where $\check{\theta}$ lies between $\hat{\theta}^{(c)}$ and $\hat{\theta}$.
Since both $\hat{\theta}^{(c)}$ and $\hat{\theta}$ are $\sqrt n$-consistent
estimators under the null hypothesis, $\check{\theta}$ is also $\sqrt n$-consistent.
By Condition (C5), Lemma 5 and the continuous mapping theorem, we have, under the null hypothesis,
\begin{center}
$-n^{-1}\frac{\partial^{2}\ell(\check{\theta})}{\partial\theta\partial\theta^\top}\xrightarrow{p}\mathcal{I}(\theta)$ and $\mathcal{I}^{-1}_{n}(\hat{\theta}^{(c)})\xrightarrow{p}\mathcal{I}^{-1}(\theta)$.
\end{center}
Accordingly, we obtain that
\[
T_{s}=\frac{1}{n}\Big\{\frac{\partial\ell(\hat{\theta}^{(c)})}{\partial\theta}\Big\}^\top\mathcal{I}_n^{-1}
(\hat{\theta}^{(c)})\Big\{\frac{\partial\ell(\hat{\theta}^{(c)})}{\partial\theta}\Big\}\]
\beq =\sqrt{n}
(\hat{\theta}^{(c)}-\hat{\theta})^\top\mathcal{I}(\theta)\sqrt{n}(\hat{\theta}^{(c)}-\hat{\theta})+o_p(1).
\eeq

By Lemma 5, we have
\begin{center}
    $n^{-1/2}\mathcal{K}^{-1/2}\frac{\partial\ell(\theta)}{\partial\theta}\xrightarrow{d}N(0,1).$
\end{center}
where $\mathcal{K}=\mathcal{K}(\theta,\mu_{3},\mu_{4})=\mathcal{I}(\theta)+\mathcal{J}(\theta,\mu_{3},\mu_{4})$ as defined in Section 2.
This, together with (S.5) and (S.6), leads to
\begin{center}
    $T_{s}=n^{-1/2}\begin{bmatrix}
    \big\{\mathcal{I}_1(\theta)-\mathcal{I}^{-1}(\theta)\big\}\frac{\partial\ell(\theta)}{\partial\theta}
    \end{bmatrix}^\top \mathcal{I}(\theta)n^{-1/2}\begin{bmatrix}\big\{\mathcal{I}_1(\theta)
    -\mathcal{I}^{-1}(\theta)\big\}\frac{\partial\ell(\theta)}{\partial\theta}\end{bmatrix}+o_p(1)$~\\
    ~\\
   $=\begin{Bmatrix}
    n^{-1/2}\mathcal{K}^{-1/2}\frac{\partial\ell(\theta)}{\partial\theta}\end{Bmatrix}^\top \mathcal{K}^{1/2}\big\{\mathcal{I}_{1}(\theta)-\mathcal{I}^{-1}(\theta)\big\}
    \mathcal{I}(\theta)\big\{\mathcal{I}_{1}(\theta)$~\\
    ~\\
    $-\mathcal{I}^{-1}(\theta)\big\}\mathcal{K}^{1/2}\begin{Bmatrix}
    n^{-1/2}\mathcal{K}^{-1/2}\frac{\partial\ell(\theta)}{\partial\theta}\end{Bmatrix}+o_{p}(1).$
    \end{center}
Using the fact that $\mathcal{I}_{1}(\theta)\mathcal{I}(\theta)\mathcal{I}_{1}(\theta)=\mathcal{I}(\theta)$ and $\big\{\mathcal{I}_{1}(\theta)-\mathcal{I}^{-1}(\theta)\big\}\mathcal{I}(\theta)\big\{\mathcal{I}_{1}(\theta)-\mathcal{I}^{-1}(\theta)\big\}=\mathcal{I}^{-1}(\theta)-\mathcal{I}_{1}(\theta)$, we further obtain that
\begin{center}
    $T_{s}=\Big\{n^{-1/2}\mathcal{K}^{-1/2}\frac{\partial\ell(\theta)}{\partial\theta}\Big\}^\top \mathcal{K}^{1/2}\big\{\mathcal{I}^{-1}(\theta)-\mathcal{I}_{1}(\theta)\big\}\mathcal{K}^{1/2}\Big\{n^{-1/2}\mathcal{K}^{-1/2}\frac{\partial\ell(\theta)}{\partial\theta}\Big\}+o_{p}(1).$
\end{center}
Subsequently, applying the continuous mapping theorem and Slutsky's theorem as well as Lemma 5, we can
demonstrate that $T_{s}$ follows a weighted chi-square distribution
$\sum_{l=1}^{p+d_{1}+d_{2}+1}\lambda_{l}(\theta,\mu_{3},\mu_{4})\chi^{2}_{l}(1)$ asymptotically,
where $\lambda_{l}(\theta)$ are the $l$-th eigenvalues of $\mathcal{K}^{1/2}\big\{\mathcal{I}^{-1}(\theta)-\mathcal{I}_{1}(\theta)\big\}\mathcal{K}^{1/2}$ for $l=1,\cdots, (p+d_{1}+d_{2}+1)$.
This completes the proof of the first part of Theorem 3.

When $\varepsilon$ is normally distributed, the matrix $\mathcal{J}_{n}(\theta,\mu_{3},\mu_{4})$ defined in Section 1 is zero.
As a result, its convergence $\mathcal{J}(\theta,\mu_{3},\mu_{4})$ defined in the Appendix of the paper
should be zero. This implies that $\mathcal{K}=\mathcal{I}(\theta)$. Using the fact that $\mathcal{K}^{1/2}\big\{\mathcal{I}^{-1}(\theta)-\mathcal{I}_{1}(\theta)\big\}\mathcal{K}^{1/2}=\mathcal{I}^{1/2}(\theta)\big\{\mathcal{I}^{-1}(\theta)-\mathcal{I}_{1}(\theta)\big\}\mathcal{I}^{1/2}(\theta)$ is symmetric and idempotent, we have
\begin{align*}
    tr\big(T_{s}\big)&=tr\Big[\mathcal{I}^{1/2}(\theta)\big\{\mathcal{I}^{-1}(\theta)-\mathcal{I}_{1}(\theta)\big\}
    \mathcal{I}^{1/2}(\theta)\Big]=tr\big\{\mathcal{I}_{p+d_{1}+d_{2}+1}-\mathcal{I}(\theta)\mathcal{I}_{1}(\theta)\big\}\\
    &=tr\begin{Bmatrix}\begin{pmatrix}
    0_{(p+2)\times(p+2)} & 0_{(p+2)\times(d_1+d_2-1)}\\
    -\mathcal{I}_{21}(\theta)\mathcal{I}_{11}^{-1}(\theta) & \mathcal{I}_{d_1+d_2-1}\end{pmatrix}\end{Bmatrix}=d_1+d_2-1.
\end{align*}
Accordingly, $T_{s}\xrightarrow{d}\chi^{2}(d_1+d_2-1)$, which completes the entire proof.

\scsubsection{Section 5: Mixture Normal Errors}

This section presents the simulation results for mixture normal errors in order to
assess the robustness of our proposed method to non-normal errors. The model simulation
setting is the same as that in Section 3 of the paper except for random errors.
Tables S1 and S2 report the results for parameter estimates and BIC selections, respectively.
They yield similar patterns to those of Tables 1 and 2 in the manuscript. In
addition, Figure S1 depicts the size and power of three tests with the inverse of probit
link function (the results for sigmoid and log-log functions are similar and thus omitted).
They yield similar findings to those obtained from Figures 2-4 of the manuscript.
In sum, our proposed estimates, selections and tests exhibit nice properties under mixture
normal errors.

\scsubsection{Section 6: An Overlapping Design}

This section considers an overlapping design.
We set $p=4$, $d_1=3$ and $d_2=4$, where the covariates of $Z$ are the same as the last three covariates of $V$ and $X$. In addition, all covariates are independent and identically generated
from the standard normal distribution $N(0,1)$ except for the intercepts  $v_{i1}=x_{i1}=1$ for $i=1,\cdots,n$. 
The associated coefficients are respectively
 $\eta=(1, 2, 3, 4)^\top$, $\alpha=(0.4, 0.6, 0.7)^\top$, $\beta=(1.2, 1.4, 1.3, 1.6)^\top$ and $\sigma^{2}=1$.
 The rest of simulation settings are the same as those in Section 3 of the paper.
Table S3 reports the results of parameter estimates, which yields a similar pattern to that of Table 1 in the manuscript and supports Theorem 1.
In sum, our proposed estimates perform satisfactorily under the overlapping design.

\newpage
\makeatletter\def\@captype{table}\makeatother

\begin{spacing}{1}
\scriptsize
\centering
\setlength{\belowcaptionskip}{0.2cm}
\begin{landscape}
{Table S1: The BIAS, SD and RMSE of the QMLEs ($\eta_{1}=1$, $\eta_{2}=2$, $\eta_{3}=3$, $\alpha_{1} =0.4$, $\alpha_{2} =0.6$, $\alpha_{3} = 0.7$, $\beta_{1}=1.2$, $\beta_{2}=1.4$, $\beta_{3}=1.3$, $\beta_{4}=1.6$ and $\sigma^{2}=1$)
under three link functions. The random errors are simulated from the  mixture normal distribution $0.9N(0,5/9)+0.1N(0,5)$.}
\label{Tab:02}
\begin{tabular}{c|cccccccccccccccc}
\hline
Link & $n$ & Measure & $\hat{\eta}_{1}$ & $\hat{\eta}_{2}$ & $\hat{\eta}_{3}$ & $\hat{\alpha}_{1}$ & $\hat{\alpha}_{2}$ & $\hat{\alpha}_{3}$ & $\hat{\gamma}$ & $\hat{\beta}_{1}$ & $\hat{\beta}_{2}$ & $\hat{\beta}_{3}$ & $\hat{\beta}_{4}$ &$\hat{\lambda}$ & $\hat{\sigma}^{2}$\\
\hline
\multirow{9}{*}{Sigmoid}&
\multirow{3}{*}{500}\\
& & BIAS &-0.001 &-0.001 &0.001 &0.000&	0.000 &0.001 &0.000 &-0.006 &-0.005 &0.006 &0.002 &-0.001 &-0.010\\
 & & SD  &0.045 &0.045 &0.045 &0.127 &0.136 &0.142 &0.002 &0.393 &0.450 &0.432 &0.490 &0.007 &0.115\\
 & & RMSE &0.045 &0.045 &0.045 &0.127 &0.136 &0.142 &0.002 &0.393 &0.450 &0.432 &0.490&	0.007 &0.115\\
 \cline{2-16}
 & \multirow{3}{*}{1000}\\
& & BIAS &0.000 &0.002&	0.000 &0.000 &0.000 &0.001 &0.000 &-0.003 &0.003 &0.003 &-0.006 &-0.001 &-0.005\\
& & SD &0.032 &0.032 &0.032 &0.089 &0.096 &0.100 &0.001 &0.272&	0.313 &0.300 &0.338 &0.003 &0.084\\
& & RMSE &0.032 &0.032 &0.032 &0.089 &0.096 &0.100 &0.001 &0.272 &0.313 &0.300 &0.339 &0.003 &0.084\\
 \cline{2-16}
& \multirow{3}{*}{2000}\\
& & BIAS &-0.001 &0.000 &0.001 &0.000 &-0.001 &-0.002 &0.000 &-0.006 &0.002 &-0.001 &-0.002 &-0.001 &-0.001\\
& & SD &0.022 &0.022 &0.022 &0.063 &0.068 &0.070 &0.000 &0.191 &0.220 &0.210 &0.238 &0.002 &0.060\\
& & RMSE &0.022 &0.022 &0.022 &0.063 &0.068 &0.070 &0.000 &0.191 &0.220 &0.210 &0.238 &0.002 &0.060\\
 \hline
 \multirow{9}{*}{Inverse of log-log}&
\multirow{3}{*}{500}\\
& & BIAS &0.000 &-0.001 &0.001 &0.001 &0.001 &0.000 &0.000 &0.005 &-0.001 &-0.002 &0.001 &0.000 &-0.010\\
 & & SD &0.045 	&0.045 &0.045 &0.081 &0.091 &0.097 &0.001 &0.382 &0.391 &0.366 &0.435 &0.004 &0.115\\
 & & RMSE &0.045 &0.045 &0.045&	0.081 &0.091 &0.097 &0.001 &0.382 &0.391 &0.366 &0.435 &0.004 &0.115\\
 \cline{2-16}
 & \multirow{3}{*}{1000}\\
& & BIAS &0.000 &0.002 &0.000 &0.000 &-0.001 &0.002 &0.000 &0.002 &0.002 &0.003 &-0.003 &0.000 &-0.005\\
& & SD &0.032 &0.032 &0.032 &0.056 &0.064 &0.069 &0.001 &0.263 &0.270 &0.256 &0.298 &0.002 &0.084\\
& & RMSE &0.032 &0.032 &0.032 &0.056 &0.064 &0.069 &0.001 &0.263 &0.270 &0.256 &0.298 &0.002 &0.084\\
 \cline{2-16}
& \multirow{3}{*}{2000}\\
& & BIAS &-0.001 &0.000 &0.001 &0.001 &-0.001 &-0.001 &0.000 &-0.005 &0.002 &0.000 &0.001 &-0.001 &-0.001\\
& & SD &0.023 &0.022 &0.022 &0.040 &0.045 &0.048 &0.000 &0.183 &0.189 &0.178 &0.210 &0.001&	0.060\\
& & RMSE &0.023 &0.022 &0.022 &0.040 &0.045 &0.048 &0.000 &0.184 &0.189 &0.178 &0.210 &0.001 &0.060\\
 \hline
 \multirow{9}{*}{Inverse of probit}&
\multirow{3}{*}{500}\\
& & BIAS &-0.001 &-0.001 &0.001 &0.000 &0.002 &0.002 &0.000 &0.002 &-0.003 &-0.001 &-0.003 &0.000 &-0.010\\
 & & SD &0.045 &0.045 &0.045 &0.088 &0.100 &0.107 &0.002 &0.379 &0.445 &0.418 &0.495 &0.006 &0.115\\
 & & RMSE &0.045 &0.045 &0.045 &0.088 &0.100 &0.107 &0.002 &0.379 &0.445 &0.418 &0.495 &0.006 &0.115\\
 \cline{2-16}
 & \multirow{3}{*}{1000}\\
& & BIAS &0.000 &0.002 &0.000 &0.000 &-0.001 &0.001 &0.000 &0.000 &0.000 &0.003 &-0.006 &0.000 &-0.005\\
 & & SD &0.032 &0.032 &0.032 &0.062 &0.070 &0.075 &0.001 &0.261 &0.308 &0.291 &0.340 &0.003 &0.084\\
 & & RMSE &0.032 &0.032 &0.032 &0.062 &0.070 &0.075 &0.001 &0.261 &0.308 &0.291 &0.340 &0.003 &0.084\\
 \cline{2-16}
& \multirow{3}{*}{2000}\\
& & BIAS &-0.001 &0.000 &0.001 &0.000 &-0.001 &0.000 &0.000 &-0.004 &0.001 &0.002 &0.000 &-0.001 &-0.001\\
& & SD &0.022 &0.022 &0.022 &0.044&	0.050 &0.053 &0.000 &0.183 &0.215 &0.203 &0.240 &0.001 &0.060\\
& & RMSE &0.022 &0.022 &0.022 &0.044 &0.050 &0.053 &0.000 &0.183 &0.215 &0.203 &0.240 &0.002 &0.060\\
 \hline
\end{tabular}
\end{landscape}
\end{spacing}

\newpage
\begin{spacing}{1}
\setlength{\belowcaptionskip}{0.2cm}
\scriptsize
\centering
{Table S2: Variable selection results  including (\%) of CF, TPR and FPR with mixture normal random errors, $0.9N(0,5/9) + 0.1N(0,5)$.}

\label{Tab:02}
\begin{tabular}{lllll|lll}
\toprule
\multirow{2}{*}{Link function} & \multirow{2}{*}{$n$} & \multicolumn{3}{l}{$\hat{S_{\alpha}}$} & \multicolumn{3}{l}{$\hat{S_{\beta}}$}\\
\cline{3-8}
~\\
& & CF & TPR & FPR & CF & TPR & FPR\\
\midrule
\multirow{3}{*}{Sigmoid}
&  500 &72.7 &90.7 &0.2 &82.3 &94.3 &0.2\\
 &  1000 &92.7 &97.6 &0.1 &99.7 &99.9 &0.0\\
 & 2000 &99.7 &99.9 &0.0 &100.0 &100.0 &0.0\\
 \hline
 \multirow{3}{*}{Inverse of log-log}
& 500 &99.0 &99.7 &0.3 &99.0 &99.8 &0.2\\
& 1000 &99.3 &100.0 &0.2 &100.0 &100.0 &0.0\\
& 2000 &100.0 &100.0 &0.0 &100.0 &100.0 &0.0\\
 \hline
\multirow{3}{*}{Inverse of probit}
& 500 &96.7 &98.9 &0.1 &98.3 &99.6 &0.2\\
 & 1000 &99.7 &100.0 &0.1 &100.0 &100.0 &0.0\\
 & 2000 &100.0 &100.0 &0.0 &100.0 &100.0 &0.0\\
 \bottomrule
\end{tabular}
\end{spacing}
~\\

\begin{figure}[H]
 \includegraphics[width=5.8in,height=2.2in]{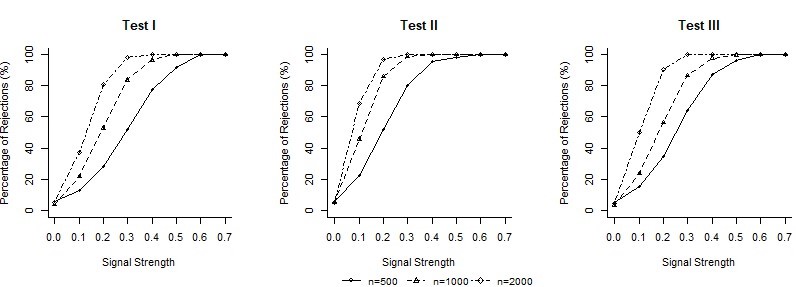}
{Figure S1: The empirical sizes and powers of Tests I--III under the inverse of
probit link function with the significance level 0.05. The signal strengths are $\delta$=0, 0.1, 0.15, 0.2, 0.25, 0.3, 0.4 and 0.7,
corresponding to the settings of $\alpha=(0.4\delta,0.6\delta,0.7\delta)^{\top}$ and $\beta = (1.2,1.4\delta,1.3\delta,1.6\delta)^{\top}$ in three tests respectively.
The random errors are independently and identically simulated from the mixture normal distribution $0.9N(0,5/9) + 0.1N(0,5)$. }
 \label{fig2}
\end{figure}

\newpage
\begin{spacing}{1}
\scriptsize
\centering
\setlength{\belowcaptionskip}{0.2cm}
\begin{landscape}
{Table S3: The BIAS, SD and RMSE of the QMLEs ($\eta_{1}=1$, $\eta_{2}=2$, $\eta_{3}=3$,
$\eta_{4}=4$, $\alpha_{1} =0.4$, $\alpha_{2} =0.6$, $\alpha_{3} = 0.7$, $\beta_{1}=1.2$, 
$\beta_{2}=1.4$, $\beta_{3}=1.3$, $\beta_{4}=1.6$ and $\sigma^{2}=1$)
under three link functions for the overlapping design. The random errors are simulated from the normal distribution $N(0,\sigma^{2})$.}
\label{Tab:01}
\begin{tabular}{c|cccccccccccccccc}
\hline
Link & $n$ & Measure & $\hat{\eta}_{1}$ & $\hat{\eta}_{2}$ & $\hat{\eta}_{3}$ & $\hat{\eta}_{4}$ & $\hat{\alpha}_{1}$ & $\hat{\alpha}_{2}$ & $\hat{\alpha}_{3}$ & $\hat{\gamma}$ & $\hat{\beta}_{1}$ & $\hat{\beta}_{2}$ & $\hat{\beta}_{3}$ & $\hat{\beta}_{4}$ &$\hat{\lambda}$ & $\hat{\sigma}^{2}$\\
\hline
\multirow{12}{*}{Sigmoid}&
\multirow{4}{*}{500}\\
& & BIAS &0.004 &0.001 &0.001 &-0.002 &0.001 &0.002 &0.002 &0.000 &-0.009 &0.008 &0.024 &0.016 &-0.003 &-0.006\\
 & & SD &0.045 &0.076 &0.075 &0.075 &0.088 &0.094 &0.096 &0.001 &0.497 &0.610 &0.544 &0.597 &0.011 &0.063\\
 & & RMSE &0.045 &0.076 &0.075 &0.075 &0.088 &0.094 &0.096 &0.001 &0.497 &0.611 &0.544 &0.598 &0.011 &0.063\\
 \cline{2-17}
 & \multirow{4}{*}{1000}\\
& & BIAS &-0.002 &-0.001 &0.001 &-0.001 &0.001 &0.001 &0.002 &0.000 &0.003 &0.010 &0.006 &0.017 &-0.001 &-0.007\\
 & & SD &0.032 &0.053 &0.053 &0.052 &0.061 &0.065 &0.067 &0.000 &0.351 &0.431 &0.377 &0.419 &0.005 &0.044\\
 & & RMSE &0.032 &0.053 &0.053 &0.052 &0.061 &0.065 &0.067 &0.000 &0.351 &0.431 &0.377 &0.419 &0.005 &0.045\\
 \cline{2-17}
& \multirow{4}{*}{2000}\\
& & BIAS &-0.001 &0.001 &-0.001 &0.000 &0.001 &0.002 &-0.001 &0.000 &0.002 &0.005 &0.013 &0.009 &0.000 &-0.004\\
& & SD &0.022 &0.038 &0.037 &0.037 &0.043 &0.046 &0.047 &0.000 &0.247 &0.304 &0.267 &0.294 &0.003 &0.031\\
& & RMSE &0.022 &0.038 &0.037 &0.037 &0.043 &0.046 &0.047 &0.000 &0.247 &0.304 &0.268 &0.294 &0.003 &0.032\\
 \hline
 \multirow{12}{*}{Inverse of log-log}&
\multirow{4}{*}{500}\\
& & BIAS &0.004 &0.000 &0.004 &-0.001 &0.001 &-0.001 &0.000 &0.000 &-0.008 &-0.003 &0.015 &0.009 &-0.002 &-0.006\\
 & & SD &0.046 &0.073 &0.071 &0.069 &0.051 &0.056 &0.058 &0.000 &0.500 &0.526 &0.452 &0.513 &0.008 &0.063\\
 & & RMSE &0.046 &0.073 &0.071 &0.069 &0.051 &0.056 &0.058 &0.000 &0.500 &0.526 &0.452 &0.513 &0.008 &0.063\\
 \cline{2-17}
 & \multirow{4}{*}{1000}\\
& & BIAS &-0.002 &-0.001 &0.002 &0.000 &0.000 &-0.001 &-0.001 &0.000 &0.005 &0.001 &0.001 &0.003 &0.000 &-0.007\\
 & & SD &0.032 &0.051 &0.050 &0.049 &0.036 &0.039 &0.040 &0.000 &0.352 &0.369 &0.312 &0.354 &0.004 &0.044\\
 & & RMSE &0.032 &0.051 &0.050 &0.049 &0.036 &0.039 &0.040 &0.000 &0.352 &0.369 &0.312 &0.354 &0.004 &0.045\\
 \cline{2-17}
& \multirow{4}{*}{2000}\\
& & BIAS &-0.001 &0.001 &-0.001 &0.000 &0.000 &0.001 &0.000 &0.000 &-0.004 &0.000 &0.004 &0.000 &-0.001 &-0.004\\
& & SD &0.023 &0.036 &0.035 &0.034 &0.025 &0.027 &0.029 &0.000 &0.245 &0.259 &0.220 &0.249 &0.002 &0.032\\
& & RMSE &0.023 &0.036 &0.035 &0.034 &0.025 &0.027 &0.029 &0.000 &0.245 &0.259 &0.220 &0.249 &0.002 &0.032\\
 \hline
 \multirow{12}{*}{Inverse of probit}&
\multirow{4}{*}{500}\\
& & BIAS &0.004 &0.000 &0.002 &0.000 &0.001 &0.000 &0.000 &0.000 &-0.008 &-0.002 &0.016 &0.011 &-0.002 &-0.006\\
 & & SD &0.045 &0.071 &0.070 &0.068 &0.053 &0.058 &0.061 &0.001 &0.527 &0.663 &0.573 &0.655 &0.013 &0.063\\
 & & RMSE &0.045 &0.071 &0.070 &0.068 &0.053 &0.058 &0.061 &0.001 &0.527 &0.663 &0.573 &0.655 &0.013 &0.063\\
 \cline{2-17}
 & \multirow{4}{*}{1000}\\
& & BIAS &-0.002 &-0.001 &0.001 &-0.001 &0.001 &0.000 &0.001 &0.000 &0.008 &-0.002 &0.003 &0.009 &0.000 &-0.007\\
 & & SD &0.032 &0.050 &0.049 &0.048 &0.037 &0.040 &0.042 &0.000 &0.373 &0.469 &0.397 &0.458 &0.006 &0.044\\
 & & RMSE &0.032 &0.050 &0.049 &0.048 &0.037 &0.040 &0.042 &0.000 &0.374 &0.469 &0.397 &0.458 &0.006 &0.045\\
 \cline{2-17}
& \multirow{4}{*}{2000}\\
& & BIAS &-0.001 &0.001 &-0.001 &0.000 &0.000 &0.001 &0.000 &0.000 &0.000 &0.002 &0.008 &0.003 &-0.001 &-0.004\\
& & SD &0.022 &0.035 &0.034 &0.034 &0.026 &0.028 &0.030 &0.000 &0.260 &0.331 &0.281 &0.319 &0.003 &0.031\\
& & RMSE &0.022 &0.035 &0.034 &0.034 &0.026 &0.028 &0.030 &0.000 &0.260 &0.331 &0.281 &0.319 &0.003 &0.032\\
 \hline
\end{tabular}
\end{landscape}
\end{spacing}

%
%
%
%